\documentclass[
nofootinbib,
 amsmath,amssymb,
 aps,
prd,
twocolumn
]{revtex4-2}
\usepackage{orcidlink}
\usepackage{graphicx}
\usepackage{dcolumn}
\usepackage{bm}
\usepackage{geometry}
\usepackage{mathtools}
\usepackage{braket}

\usepackage{algorithm2e}
\usepackage{lineno}

\usepackage [english]{babel}
\begin{document}
\newcommand{\nux}{\nu_{x}}
\newcommand{\nue}{\nu_{e}}
\newcommand{\anue}{\overline{\nu}_{e}}
\newcommand{\fac}{\langle \sigma \rangle}
\newcommand{\sigmafmat}{\boldsymbol{\sigma_f}(E_\nu)}
\newcommand{\imotelltalefigscale}{0.45}
\newcommand{\placeholder}{\textbf{PLACEHOLDER}}
\newcommand{\sumnu}{\sum m_\nu}

\preprint{APS/123-QED}

\title{Investigating the Neutrino Mass Ordering Problem via Ternary Plots}

\author{Alexander Migala
\orcidlink{0009-0003-7860-3805}}
\email{amigala@ucsd.edu}
 \altaffiliation[Now at ]{Department of Physics, University of California, San Diego, La Jolla, CA 92093, USA}
\affiliation{Department of Physics, Duke University, Durham, NC 27708, USA}
\author{Kate Scholberg
\orcidlink{0000-0002-7007-2021}}
\affiliation{Department of Physics, Duke University, Durham, NC, 27708, USA}
\date{\today}

\begin{abstract}
We explore what may be deduced about the neutrino mass ordering problem from the observation of core-collapse supernova burst neutrinos in modern terrestrial detectors. We employ ternary plots in a novel way to visualize the time evolution of the flavor composition of various supernova neutrino flux models from the SNEWPY software package. Through our analysis of several models using a simplified unfolding process, we have explored potential robust discriminants between the normal and inverted mass orderings. We find that the normal and inverted mass orderings tend to occupy different regions in ternary space across different models.
\end{abstract}
\maketitle

\section{\label{sec:level1} Introduction}
In the current three-flavor paradigm of massive neutrinos, flavor eigenstates are characterized as a superposition of mass eigenstates, and the weakly interacting flavor eigenstates, $\nu_e$, $\nu_\tau$, and $\nu_\mu$, are connected to the mass eigenstates $m_1$, $m_2$, and $m_3$, by the PMNS (Pontecorvo-Maki-Nakawaga-Sakata) matrix $\ket{\nu_f} = \sum_{i=1}^{N} U_{fi}^* \ket{\nu_i}$, where

\begin{equation}
\begin{split}
&U = 
\left(
\begin{array}{c c c}
1 & 0 & 0\\ 
0 & c_{23} & s_{23} \\  
0 & -s_{23} & c_{23} 
\end{array} \right)... \\
& \left(
\begin{array}{ccc}
c_{13} & 0 & s_{13}e^{-i\delta}\\ 
0 & 1 & 0 \\  
-s_{13}e^{i\delta} & 0 & c_{13} 
\end{array} \right)
\left(
\begin{array}{ccc}
c_{12} & s_{12} & 0\\ 
-s_{12} & c_{12} & 0 \\  
0 & 0 & 1 
\end{array} \right).
\label{eqn:mns}
\end{split}
\end{equation}

In this picture, neutrinos oscillate in flavor as they propagate in vacuum. The relevant PMNS matrix mixing angle parameters  are $\theta_{23}$, $\theta_{12}$, and $\theta_{13}$; $c_{ij}$ and $s_{ij}$ are the cosine and sine of $\theta_{ij}$, respectively. $\delta$ is a complex phase that is sensitive to charge-parity violation in oscillation.
The full phenomenology of three-flavor neutrino oscillation depends also on the mass-squared differences, $\Delta m^2_{ij} \equiv m_i^2-m_j^2$, in addition to the PMNS matrix angles.  In the presence of  matter, the  Mikheyev-Smirnov-Wolfenstein (MSW) effect results in a matter-density-dependent modulation of the oscillation probability.

The absolute neutrino mass scale does not affect oscillation phenomenology.  However, the ordering of the three masses does. As it is conventional, we take $m_2>m_1$ in the smaller-magnitude  $\Delta m^2_{21}$ mass-squared difference measured in solar-neutrino oscillations. The larger-magnitude mass-squared difference, $\Delta m^2_{32}$, has unknown sign.
If $\Delta m^2_{32}>0$, and $m_1 < m_2 << m_3$, then the mass ordering (MO) is known as \textbf{normal} (NMO), and we have two light and one heavy mass states.   If $\Delta m^2_{32}<0$, and $m_3 << m_1 < m_2$, then the mass ordering is \textbf{inverted} (IMO) and we have two heavy and one light mass states.

Over the past few decades, neutrino oscillation experiments have yielded values for mass-squared differences and PMNS mixing angles \cite{pdg_neutrino_review}.  Current best-fit values from the Particle Data Group~\cite{pdg_neutrino_review} are:
\begin{equation}
\begin{split}
    & \text{Inverted:}~\Delta m^2_{32}=(-2.529\pm 0.029)\times 10^{-3}~\text{eV}^2 \\
    & \text{Normal:}~\Delta m^2_{32}=(2.455\pm 0.028)\times 10^{-3}~\text{eV}^2
\end{split}    
\end{equation}
The best-fit values of the magnitude of $\Delta m^3_{32}$ are slightly different for the normal and inverted cases because of matter effects on long-baseline oscillations.

The mass ordering is one of the remaining major unknowns of the three-flavor oscillation picture. In addition to being of intrinsic interest for neutrino mass models, 
the solution to the mass-ordering problem has immediate impact on several areas of neutrino physics.
An important example is neutrinoless double-beta decay, a rare type of nuclear decay that, if observed, indicates that neutrinos are their own antiparticles.   The mass ordering determines the sensitivity regions of the corresponding experiments~\cite{pdg_neutrino_review}.

Neutrino mass ordering has an indirect but important impact on cosmology through its relation to the minimum allowed sum of neutrino masses. In the early universe, neutrinos were in thermal equilibrium with a plasma of electrons, positrons, and photons~\cite{kolb_edward_w_early_1990}. However, during the weak decoupling epoch, the interaction rate of neutrinos fell below the expansion rate of the Universe. At that time, neutrinos were highly relativistic, so their masses had a negligible effect on the energy density. After decoupling, neutrinos freely propagated, behaving as radiation at early times and later as potentially non-relativistic matter. This transition affected both the expansion history and the growth of structure, leaving imprints on observables such as the cosmic microwave background and large-scale structure~\cite{pdg_neutrino_review}.  Cosmological surveys are primarily sensitive to the sum of neutrino masses, $\sum m_\nu$. For example, recent results from the Dark Energy Spectroscopic Instrument (DESI), in combination with other datasets, can be interpreted as constraints on $\sum m_\nu$ within the standard $\Lambda$CDM framework~\cite{collaboration_desi_2025}. Oscillation measurements imply lower bounds of approximately 0.059 eV for the NMO and 0.10 eV for the IMO. However, recent cosmological fits that prefer low values of $\sum m_\nu$ are in mild tension with these limits; for example, Ref.~\cite{elbers_constraints_2025} finds $\sum m_\nu < 0.053$ eV (95\%) after accounting for the physical boundary. Extensions to $\Lambda$CDM, such as a dynamical dark-energy $w_0w_a$CDM model, can alleviate this tension~\cite{elbers_constraints_2025}. Because of the systematic uncertainty in cosmological modeling, it is currently difficult to say with confidence that cosmological results will constrain the mass ordering. Rather, independent mass-ordering determination will provide important input to cosmological fits.

The worldwide program to fully characterize three-flavor oscillation parameters is approaching the mass ordering in two major ways.  One approach is via the matter effects in long-baseline experiments, where the neutrinos are either naturally occurring neutrinos from the atmosphere or produced artificially in beams.
The results from  T2K, Super-Kamiokande, and NOvA on mass ordering are so far inconclusive \cite{t2k_sk_abe_first_2025,T2K:2025wet}. Next-generation long-baseline beam experiments include DUNE (Deep Underground Neutrino Experiment) \cite{dune_abi_prospects_2021} and Hyper-Kamiokande \cite{collaboration_sensitivity_2025}; they will be complemented by atmospheric neutrino measurements in the same detectors, as well as in large arrays such as IceCube~\cite{IceCube:2025chb} and KM3NeT~\cite{KM3NeT:2021ozk}. DUNE, with its 1300~km baseline, has especially good sensitivity to mass ordering.
JUNO~\cite{abusleme_potential_2025}, a 20-kton scintillator experiment observing reactor $\bar{\nu}_e$, takes a different approach to mass ordering by looking for the subtle oscillation pattern changes from the different mass-squared differences. Overall, if the worldwide program proceeds as planned, it is reasonable to expect that we will know the mass ordering from terrestrial experiments by the mid 2030's.

While laboratory experiments may well have success in determining the mass ordering in the next decade, it is also possible that a future observed core-collapse supernova event may answer the neutrino mass ordering problem.

As explained in more detail in section \ref{sec:supernova_neutrino_signal}, core-collapse supernovae have a distinct neutrino signal with information about the mass ordering imprinted on the time, energy, and flavor structure of the neutrino burst due to flavor transformation effects that depend on the MO~\cite{scholberg_supernova_2017}. These signatures are measurable in terrestrial detectors~\cite{ikeda_search_2007, simpson_sensitivity_2019, abe_real-time_2016, collaboration_icecube_2015}. The observable mass-ordering signatures are, however, complicated by uncertainties in the flux models and neutrino flavor transformation processes, making a robust, model-independent signature difficult. We note that if mass ordering is already known from oscillation experiments at the time of a supernova burst detection, it will be possible to better constrain models of astrophysical and oscillation processes of the supernovae. 

The goal of this paper is to explore the predicted detected signals for a range of core-collapse supernovae flux models and visualize the results to illuminate where a robust discriminant could lie between NMO and IMO. In particular, we explore the use of ternary diagrams to visualize the reconstructed flavor evolution for different mass-ordering assumptions; we refer the reader to Ref.~\cite{ternary_diagram, Fogli:1995uu} for more information about ternary diagrams. While ternary diagrams have been used in the context of analyzing neutrino flavor composition \cite{abbasi_detection_2022,quigg_cosmic_2008, Fogli:1995uu, Fogli:1996ne, Fogli:2009rd, Capanema:2024hdm, Capanema:2025htx}, this paper specifically considers a ternary track visualization of flavor composition changing in time over the course of a supernova neutrino burst.

Section~\ref{sec:supernova_neutrino_signal} describes the neutrino signal from a supernova; Sec.~\ref{sec:mo_signatures} discusses mass ordering signatures visible in the flavor transformation and our ternary diagram visualization; Sec~\ref{sec:neutrino_detection} describes detection of neutrinos and observable interaction channels; Sec.~\ref{sec:unfolding} discusses our simple unfolding method; Sec.~\ref{sec:models} describes results for different model families and finally, section \ref{sec:discussion} summarizes the results.

\section{The Supernova Neutrino Signal \label{sec:supernova_neutrino_signal}}
At the end of a massive star's lifetime, a core-collapse event ensues that results in the formation of a compact remnant: either a neutron star or a black hole. In the former case, a bright optical supernova may result; in either case, the enormous binding energy of the compact remnant is emitted in the form of a bright burst of neutrinos over an interval of a few tens of seconds.   Modern models of neutrino production from core collapse (consistent with the low-statistics 1987A observation of neutrinos~\cite{1987a, hirata_observation_1988, bionta_observation_1987, alexeyev_detection_1988}) tend to agree on the main features of the neutrino signal.

\subsection{Expected Neutrino Burst Fluxes}
\begin{figure*}
    \centering
    \includegraphics[scale=0.55]{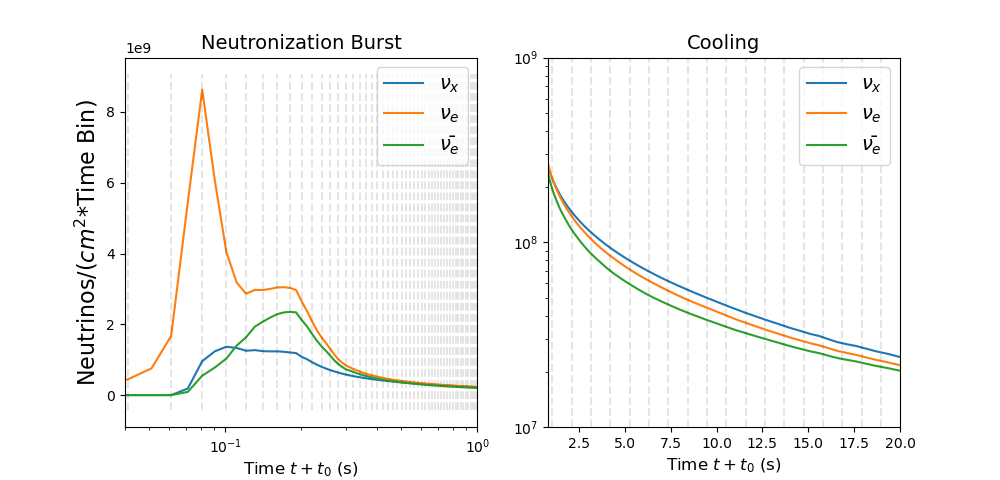}
    \caption{Supernova neutrino flux evolution in the Nakazato index 0 model (see App.~\ref{sec:model_descriptions_detailed})~\cite{nakazato_supernova_2013}, showing generic features. Left: burst phase with neutronization spike. Right: the cooling phase.  Time bins are uniformly-sized and offset by the model's absolute initial time plus $1~\text{ms}$. Vertical lines show the spacing of every two time bins and every 100 bins on the left and right, respectively. Each time bin is 0.01 seconds. For effective visualization, the $\nux$ channel flux has been divided by four, as it is a combination of $\nu_\tau$, $\bar{\nu}_\tau$, $\nu_\mu$, and $\bar{\nu}_\mu$.}
    \label{fig:supernova_formation}
\end{figure*}

Fig. \ref{fig:supernova_formation} shows an example simulated model of core-collapse supernova neutrino fluxes from the Nakazato suite \cite{nakazato_supernova_2013}.   
Note that in this figure, and throughout the rest of this work, $\nux=\nu_\tau+\nu_\mu +\bar{\nu}_\tau+\bar{\nu}_\mu$; this combination is made since the production and interaction physics of the muon and tau flavors are very similar in the below-charged-current-threshold regime of relevance here.
From the two example no-flavor-transition plots in Fig. \ref{fig:supernova_formation}, it can be seen that the ratio of each neutrino flavor evolves with time. This time dependence is explained by a few key processes~\cite{mirizzi_supernova_2016}.

The characteristic $\nue$ spike, the ``neutronization burst," seen in the left panel of Fig.~\ref{fig:supernova_formation}, is explained by electron capture, where electrons and protons interact to become neutrons and electron neutrinos, during the initial infall of the star \cite{mirizzi_supernova_2016}.  These later escape as the shock passes the neutrinosphere.

After the neutronization burst, there is an explosion and accretion phase in which non-electron flavors turn on and the neutrinos' energy, flavor, and time structure can be modulated by shock waves, standing accretion shock instability, and other astrophysical effects.

Finally, as the proto-neutron star cools, neutrinos continue to release over some tens of seconds, leaving the deleptonized compact remnant behind \cite{mirizzi_supernova_2016}.

\subsubsection{Neutrino Energy Spectra}

At any given moment in time, the neutrino spectrum for each flavor is well described by a three-parameter expression, known as the ``pinched-thermal" or ``Garching" parameterization~\cite{Tamborra:2012ac}, 

\begin{equation}
\begin{split}
       \phi(E_{\nu}) =& N_0 \frac{(\alpha+1)^{\alpha+1}}{\langle E_\nu \rangle \Gamma(\alpha+1)}
       \left(\frac{E_{\nu}}{\langle E_{\nu} \rangle}\right)^{\alpha}\times \\ &\exp\left[-\left(\alpha + 1\right)\frac{E_{\nu}}{\langle E_{\nu} \rangle}\right]
\end{split}
\label{eqn:pinched}
\end{equation}

In this expression, $E_\nu$ is the neutrino energy and $\alpha$ is the ``pinching" parameter that determines the width of the tails of the distribution.  $N_0$ is a normalization which depends on the total luminosity.

The supernova burst flux evolution can therefore be described as a function of time for each of the three effective flavors by the specification of nine parameters as a function of time.  We use this formulation in order to describe the time dependence of the predicted signal for the given models.

\section{Flavor Transformations and Mass Ordering Signatures \label{sec:mo_signatures}}

Over the evolution of the supernova, neutrino flavors transform due to oscillation effects in the presence of matter. Interactions of neutrinos with themselves can also have non-negligible effects in phases when the neutrino number density is such that self-interaction potential exceeds the matter potential.  Flavor transformations can be complex and are at the time of this writing not yet fully understood. Ref.~\cite{johns-richers} reviews the current status of theoretical understanding.  In this study, we make the simplified assumption that the matter potential is dominant over the full evolution of the supernova and only apply the adiabatic MSW transformation assumption~\cite{scholberg_supernova_2017}, which is MO-dependent.   In the adiabatic MSW assumption, the fluxes exiting the supernova can be expressed as:
\begin{eqnarray}
F_{\nu_e} &=& F^0_{\nu_x} \,\ \,\ \,\ \,\ \,\ \,\ \,\ \,\  \,\ \,\ \,\ \,\   \,\ \,\  \,\ \,\ \,\ \,\ \,\ \textrm{(NMO)}, \label{eq:msw_nmo}\\
F_{\nu_e} &=&  \sin^2 \theta_{12} F^0_{\nu_e} +
\cos^2 \theta_{12} F^0_{\nu_x} \,\ \textrm{(IMO)} \,\,
\label{eq:msw_imo}
\end{eqnarray} and
\begin{eqnarray}  
F_{\bar\nu_e} &=& \cos^2 \theta_{12} F^0_{\bar\nu_e} + \sin^2 \theta_{12} F^0_{\bar\nu_x}   \,\ \textrm{(NMO)}, \label{eq:msw_nmo_anti}\\
F_{\bar\nu_e} &=&   F^0_{\bar\nu_x}  \,\,\,\,\,\,\,\,\,\,\,\,\,\,\,\,\,\,\,\,\,\,\,\,\,\,\,\,\,\,\,\,\,\,\,\,\,\,\,\,\,\,\,\,\,\,\,\,\,\,\,\,\,\,\,\,\,\, \textrm{(IMO)} \label{eq:msw_imo_anti}
\end{eqnarray} 

where $F_{\nu_f}$ is the flux of flavor $f$ (with $f=x$ representing any of muon and tau neutrinos or respective antineutrinos).
Under this assumption, for NMO, $\nue$ spectra will be substituted for the typically tharder $\nux$ spectra, whereas $\bar{\nu}_e$ are only partially transformed. For IMO, the full flavor swap occurs for antineutrinos instead.

\subsection{Ternary Diagram Visualization of Flavor Evolution}

We seek a discriminant between the two orderings in the time and flavor dependence of the neutrino fluxes that is robust against supernova model assumptions. We employ ternary diagrams to aid in visualization of the evolution of flavor content as a function of time during the burst in order to identify mass-ordering-dependent patterns.

Each point on the ternary diagram is a tuple of fractions that add to unity; in our graphs, we scale the fractions by 100. This means that each point characterizes the fraction of each item at a given time. As the flavor content of the supernova burst flux evolves, a track is traced out on the ternary diagram.  Since flavor content vs time depends on mass ordering, the idea is that the features of the observed track in ternary space will help to discriminate the mass ordering.  We define specific fractional quantities to enhance visualization of mass-ordering-dependent features in the evolution.

To create these ternary diagrams, this work makes use of the SNEWPY \cite{snewpy} and SNOwGLoBES software packages \cite{snowglobes}. The SNEWPY  (Supernova Neutrino Early Warning Models for Python) package is primarily used as a database of the core-collapse neutrino fluxes produced from different supernova models observed at Earth.  SNOwGLoBES is the software package that simulates detector responses of neutrino sources given a flux and the detector characteristics. SNEWPY provides the software wrappers for several supernova model families, and SNOwGLoBES simulates the event rates and fluxes. Our package, \texttt{snewpyternary}, connects these two packages together and adds additional functionality to produce the ternary diagrams.   

\begin{figure*}
    \centering
    \includegraphics[width=0.75\linewidth]{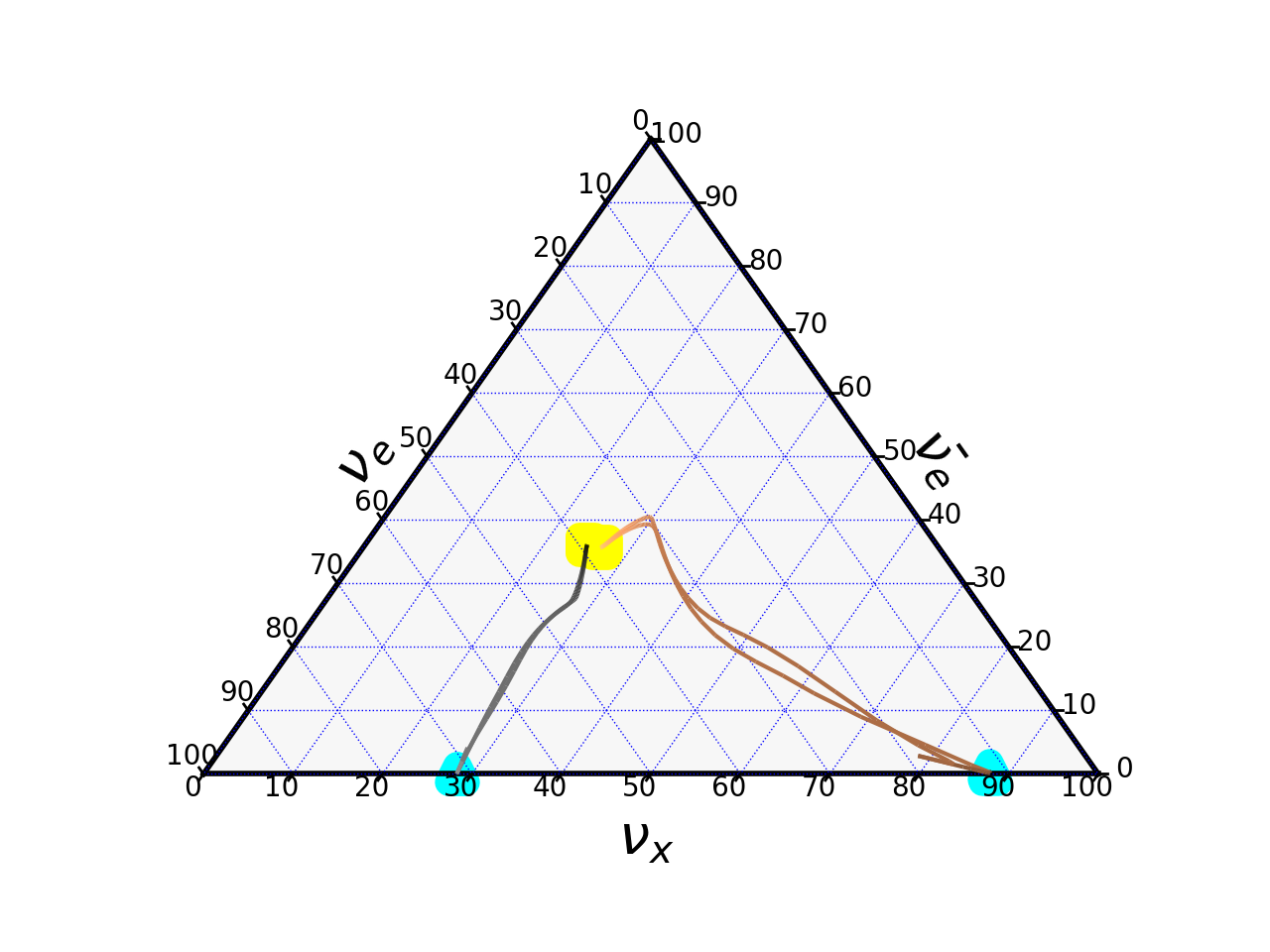}
    \caption{Ternary diagram for the truth flux of Nakazato index 0 and 2 (see App.~\ref{sec:model_descriptions_detailed}) under NMO/IMO adiabatic prescription. This is cumulatively summed over time. For effective visualization, $\nux$ was divided by six before calculating the ternary fraction, so that the trajectory ends near the center of the diagram. Time bins are logarithmically-separated. The time evolution for IMO is shown in black and white while NMO is shown in copper. Darker colors for the copper track correspond to earlier times. The lighter colors for the gray track correspond to earlier times. The start and end of the time evolution are shown in cyan and yellow, respectively.}
    \label{fig:nakazato_s0_truth_flux_csum}
\end{figure*}

\begin{figure*}
    \centering
    \includegraphics[width=1\linewidth]{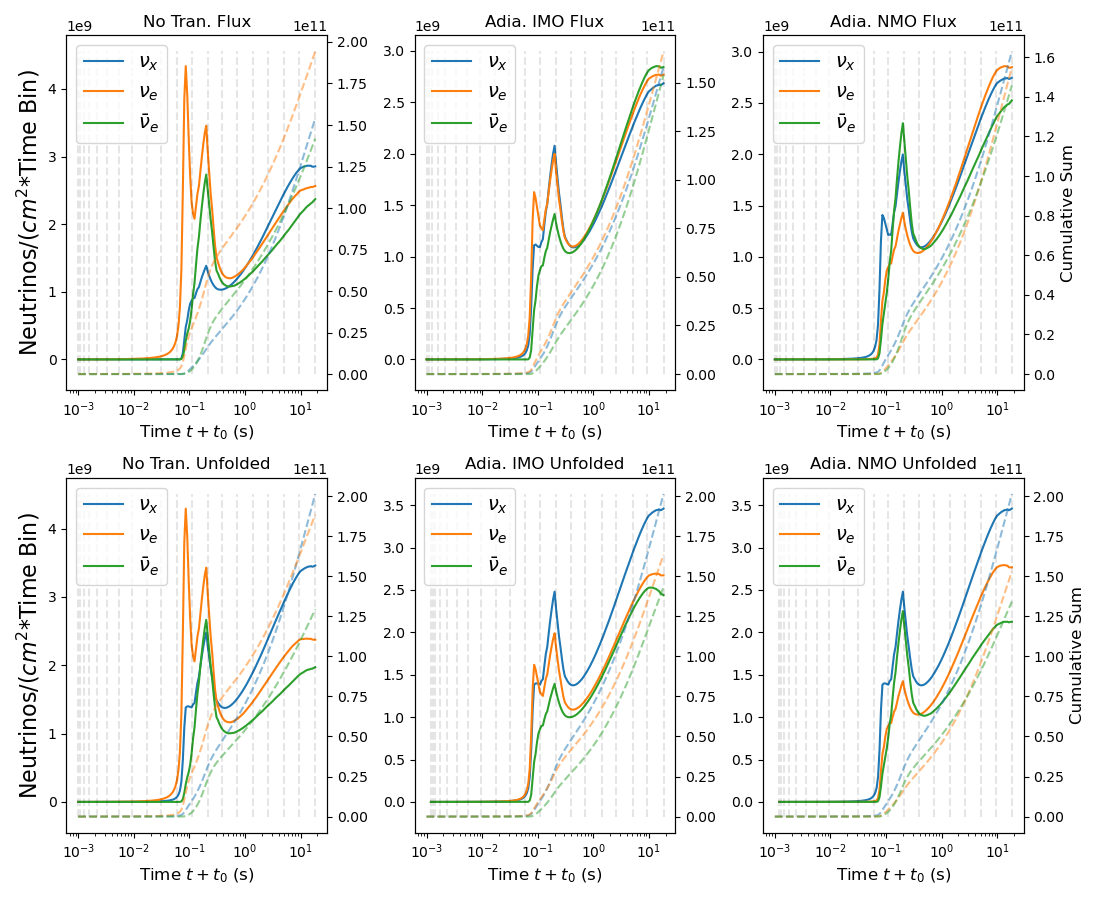}
    \caption{Flux profiles for the Nakazato s0 model under different oscillation prescriptions. The time axis is shown as an offset of the model's absolute initial time plus $1~\text{ms}$. Top row: truth flux from the model. Bottom row: simple-unfolded counterpart of truth flux. Left: no transformation applied with logarithmically sized time bins. Center: Adiabatic IMO prescription applied with logarithmically sized bins. Right: Adiabatic NMO prescription applied with logarithmically sized bins. In each of the plots, a vertical line is placed every 10 time bins. For effective visualization, the $\nux$ channel flux has been divided by four, as it is a combination of $\nu_\tau$, $\bar{\nu}_\tau$, $\nu_\mu$, and $\bar{\nu}_\mu$. The dashed lines and right axis show the cumulative sum.}
    \label{fig:flux_comparison}
\end{figure*}

We show an example of a flux ternary diagram for a single model (submodel index 0 in Appendix~\ref{sec:model_descriptions_detailed}) from the Nakazato 2013 family of models \cite{nakazato_supernova_2013}. Nakazato index 0 fixes the progenitor's mass, equation of state (EOS), revival time, and metallicity. The specific values are $20~M_\odot$, Shen's relativistic nuclear EOS~\cite{shen_paper_a, shen_paper_b}, 100 ms, and 0.004, respectively. See Appendix~\ref{sec:model_descriptions_detailed} for more details. Each model in SNEWPY's collection includes the simulation's luminosity spectra as a function of flavor, energy, and time. From this luminosity, we can calculate the total fluence in each time slice. For a specific model, this produces the fluence per time bin vs time for each flavor (see Fig.~\ref{fig:flux_comparison}). From this, we may immediately produce ternary diagrams for the IMO and NMO cases, shown in Fig.~\ref{fig:nakazato_s0_truth_flux_csum}.

Figure~\ref{fig:nakazato_s0_truth_flux_csum} shows an example of trajectories in ternary space for this model for the case of NMO and IMO. There is evidently a different trajectory in ternary space for the truth flux flavor composition, dependent on mass ordering. We next explore the observable signal.

\section{\label{sec:neutrino_detection} Neutrino Detection}

The burst of supernova neutrinos can be observed in terrestrial detectors via charged-current (CC) and neutral-current (NC) interactions with matter.  Detection techniques are reviewed in Ref.~\cite{scholberg_supernova_2012}.  CC weak interactions depend on the flavor content of the neutrino at arrival, whereas NC interactions, to a very good approximation, take place independent of flavor content.  CC thresholds for muon and tau flavor are about 110~MeV and 3.5~GeV, respectively; therefore, only $\nu_e$ and $\bar{\nu}_e$ components of the supernova flux are accessible via CC interactions as the vast majority of the neutrinos from the supernova have energies well below 100~MeV.  Therefore, $\nux$ are experimentally accessible only via NC interactions. The $\nux$ flavors are also produced at similar rates in the supernova.  For this reason, they are considered together as $\nu_x$.

The primary interaction channels in existing and near-future detectors are inverse beta decay on protons (IBD), $\nu_e$ charged-current on argon ($\nu_e$CC), and elastic scattering on electrons (eES).  We consider also a potentially taggable NC channel available in scintillator detectors, the excitation of $^{12}$C, to produce a 15-MeV deexcitation gamma ray.

The IBD interaction, 

\begin{equation}
    \anue + p \rightarrow n + e^+
    \label{eqn:ibd_simplified}
\end{equation}

is dominant in detectors with a high fraction of free protons, including water and scintillator detectors.
The final-state positron produces Cherenkov radiation in water or scintillation photons in liquid scintillator, observable with photomultiplier tubes. The IBD interaction is also taggable via delayed capture of the neutron on protons or a dopant such as gadolinium~\cite{shinoki_measurement_2023,abe_first_2022,abe_second_2024}. 

The CC interaction of $\nu_e$ and argon nuclei, 
\begin{equation}
    \nue + \prescript{40}{}{\rm Ar} \rightarrow e^- + \prescript{40}{}{\rm K}^*
\end{equation}
is dominant in liquid-argon detectors.  Elastic scattering on electrons, which occurs in all detectors, is primarily sensitive to $\nu_e$, although the eES interaction sample will contain some $\bar{\nu}_e$ and $\nu_x$.

 These interaction channels are measured in different types of real-world detectors. We can combine the calculated event rates for different channels in various detectors to evaluate the flavor content at Earth.
 For our analysis, we estimated the supernova burst event rates in a 200 kt water Cherenkov detector, a 40 kt liquid argon detector, and a 20 kt scintillator. These were selected to be characteristic of upcoming experiments. The water detector, liquid argon detector, are characteristic of Hyper-K, DUNE, and JUNO, respectively.

\begin{figure*}
    \centering
    \includegraphics[width=1\linewidth]{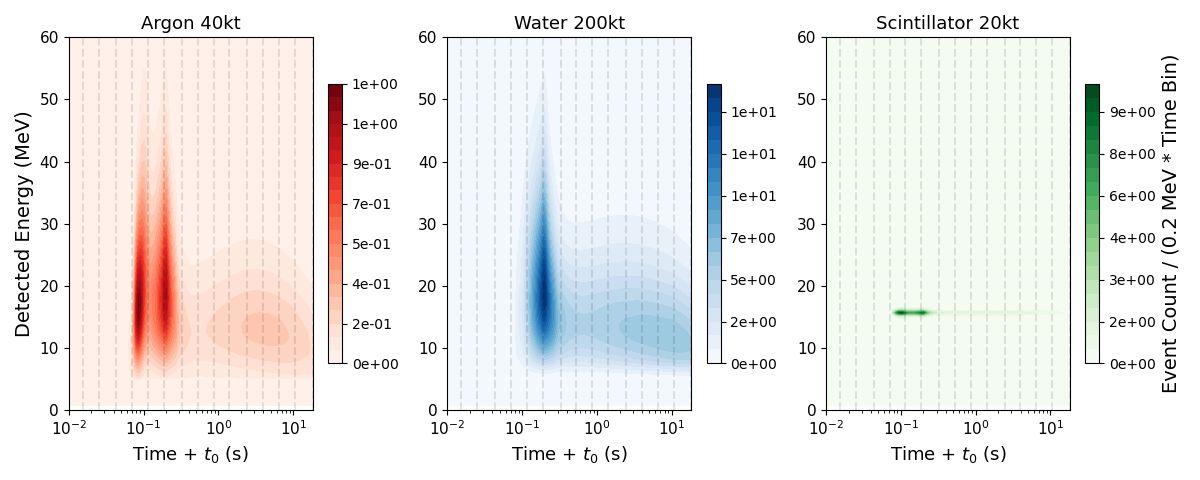}
    \caption{The computed supernova signal as seen in different detection technologies for a Nakazato flux model (index 0 in App.~\ref{sec:model_descriptions_detailed})~\cite{nakazato_supernova_2013}. Left: detected  $\nu_e$CC events vs energy and time for a generic argon 40-kiloton detector. Center: detected events for the IBD channel for a generic 200-kiloton water detector. Right: detected events for the NC channel for a generic 20-kiloton scintillator. $t_0$ denotes the absolute start time of the flux model plus $0.1~\text{ms}$. Each time bin (gray vertical lines) is 0.09 seconds.}
    \label{fig:spectra_signal_figure}
\end{figure*}

Ideally, detection of neutrino interactions in the burst would preserve all details of the flavor, energy and time structure of the burst.  In practice, detectors observe specific interaction channels according to their detection materials, and can only observe the particles in the final state.  Additionally, detectors cannot perfectly reconstruct the incoming neutrino energy from the final-state products of the interaction.  Observable final-state products may not retain all of the neutrino energy, either because the scattered neutrino takes away some of the original available energy (as in e.g., eES events) or because energies of some of the final-state products are not easily observable  In addition, intrinsic detector resolution smears observed energies.  We approximate both final-state product intrinsic energy distributions and detector effects using SNOwGLoBES for our detector configurations; example results can be seen in Fig.~\ref{fig:spectra_signal_figure}.

For CC interactions on nuclei, the final-state leptons tend to retain most of the energy of the neutrino, which is observable with reasonable resolution in many detectors.  For scattering interactions, energy is lost; however, the recoil energy distribution is well understood so that information on the neutrino energy spectrum can be determined statistically.  In general, practical detector time resolutions are good, typically sub-ms, so we assume this will not be a significant source of uncertainty.   The flavor of the incoming neutrino, in the three categories, can be obtained via channel tagging algorithms.  Here we focus on flavor content as a function of time, assuming that interaction times can be determined.

While each detector type will observe a multi-channel, multi-flavor signal, we simplify our analysis by considering only specific dominant or taggable channels as  ``proxy channels" for each neutrino flavor component of the flux.  We make the additional simplifying assumption that interactions can be selected with perfect efficiency.  The $\nu_e$CC channel in the representative argon detector is used as the $\nue$ proxy; the IBD channel in our representative scintillator and water detectors is used as the $\bar{\nu}_e$ proxy;  and the NC 15-MeV gamma channel in the representative scintillator detector is used as the $\nux$ proxy.\footnote{We note that the IBD events for the scintillator could be added to the total IBD signal; we did not do this for simplicity.} See Tab.~\ref{tab:best_channel_proxy_configuration}.

\begin{table}
\begin{center}
    \textbf{Interaction Channels for Flavor Proxies}
    \begin{tabular}{|| c c c c ||}
    \hline
    Detector & $\nu_x$  & $\nu_e$  & $\overline{\nu}_e$ \\
    \hline
    Scint & NC & $\nu_e$CC $^{12}${}C & IBD \\
    & & $\nu_{e}$CC $^{13}${}C & \\
    Argon & NC & $\nu_{e}$CC $^{40}${}Ar & \\
    Water & NC & $\nu_{e}$CC $^{16}${}O & IBD \\
    \hline 
    \end{tabular}
\end{center}
\label{tab:proxy_schema_all_channels}
\caption{Interaction channels available in each detector type as flavor proxies.}
\end{table}

\begin{table}
\begin{center}
    \textbf{Best Channel Proxy Configuration}\\
    \begin{tabular}{|| c c c c ||}
    \hline
    Detector & $\nu_x$  & $\nu_e$  & $\overline{\nu}_e$  \\
    \hline
    Scint & NC & - & - \\
    Argon & - & $\nu_e$CC & - \\
    Water & - & - & IBD \\
    \hline
    \end{tabular}
\end{center}
\caption{Dominant channels in each detector type chosen as flavor component proxies.}
\label{tab:best_channel_proxy_configuration}
\end{table}

\begin{figure*}
    \centering
    \includegraphics[width=1\linewidth]{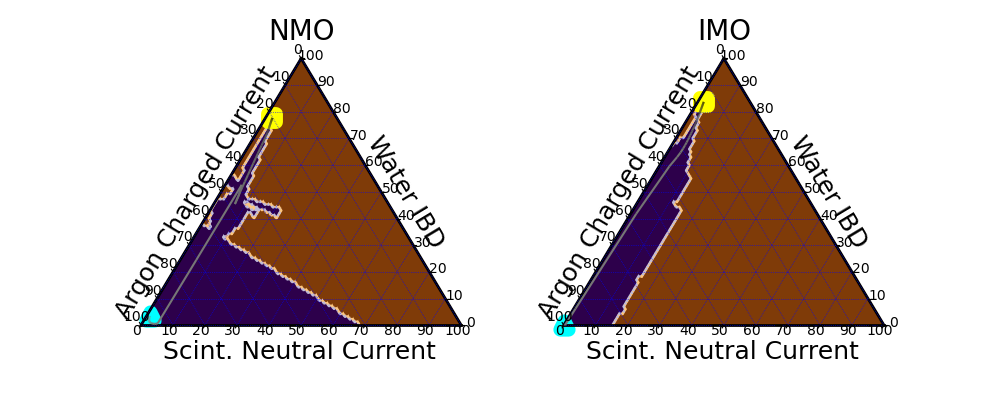}
    \caption{Nakazato submodel 0 channel event rate evolution for both IMO (right) and NMO (left); axes are labeled according to the channel in the best channel proxy configuration in Tab.~\ref{tab:best_channel_proxy_configuration}. The time evolution is shown in black and white; lighter colors correspond to earlier times. The start and end of the time evolution are shown in cyan and yellow, respectively.  The regions are colored according to the prescription in Appendix~\ref{sec:estimated_uncertainties}.}
    \label{fig:nakazato_0_ndet}
\end{figure*}

We can create a ternary diagram based solely on the event rates for each proxy channel.   
Fig.~\ref{fig:spectra_signal_figure} shows an example of a detector response from a neutrino signal. In each time bin in ternary space, we calculate the respective raw event count fractions for $\nux$, $\anue$, and $\nue$ proxies and plot this as a point (see Eq. \ref{eqn:fractionaltform}). Nakazato model 0's raw detector count (denoted $N_{\rm det}$) evolution in ternary space is shown in Fig.~\ref{fig:nakazato_0_ndet}.  The colored regions correspond to 1-sigma statistical regions determined according to Appendix~\ref{sec:estimated_uncertainties}.

Plotting the raw detected event-rate evolution on a ternary plot did not reveal any robust qualitative discriminants between NMO and IMO. The tracks for other model families were too variable between IMO and NMO to immediately discern a robust discriminant.

It is not surprising that the mass ordering discrimination pattern is not easily discernible, given that event counts depend strongly on the interaction rates for the different detector target masses and cross sections. To retrieve a visualization of the ternary trajectory pattern for different orderings, we perform a simple unfolding to estimate the truth fluxes from the event counts, as detailed in the next section.

\begin{figure*}

\begin{tabular}{c|c}
     IMO (Nakazato 2013) & NMO (Nakazato 2013) \\
     \hline \hline \\
     \includegraphics[scale=\imotelltalefigscale,trim={0 0 0 1.39cm},clip]{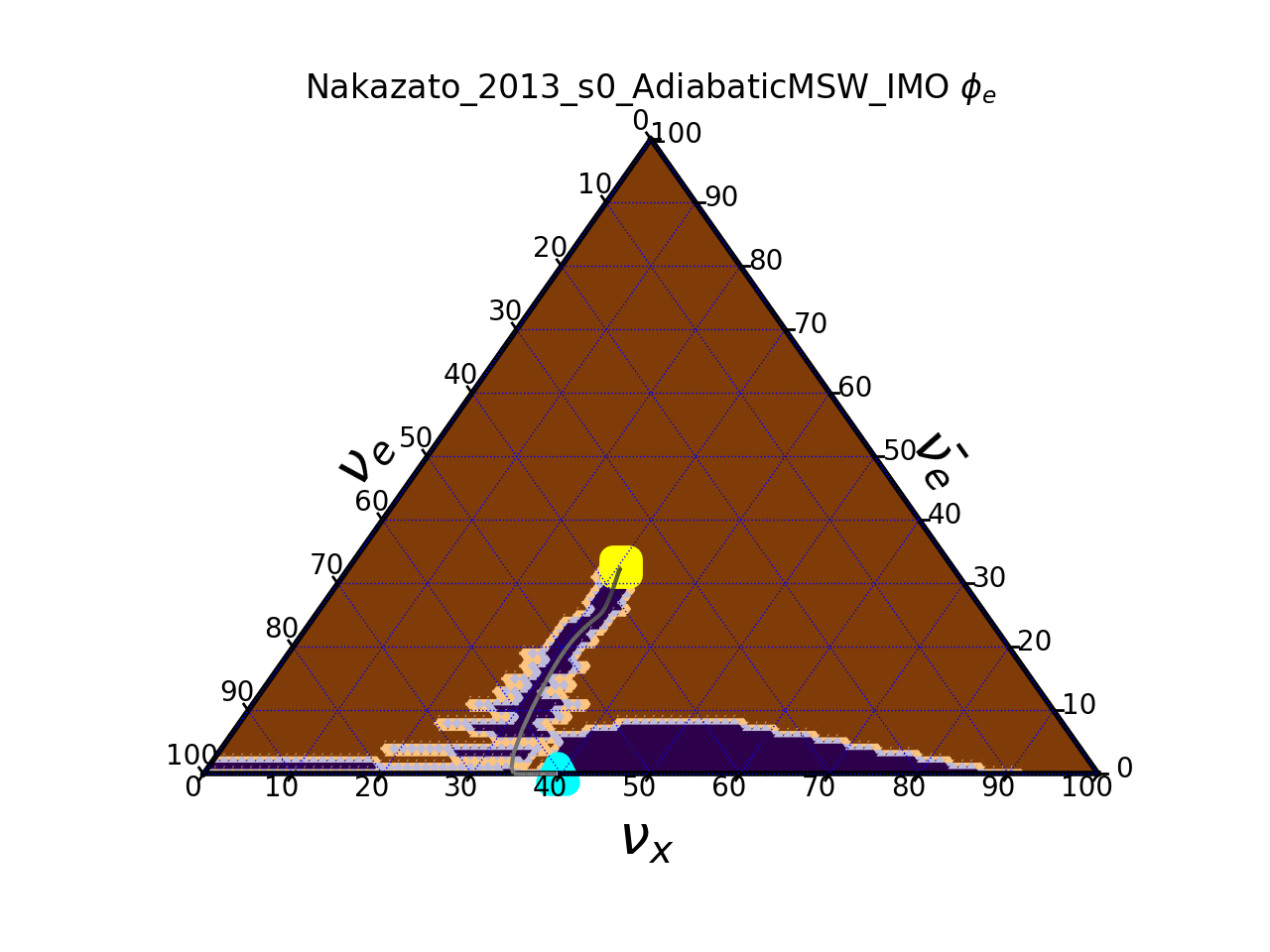} & \includegraphics[scale=\imotelltalefigscale,trim={0 0 0 1.39cm},clip]{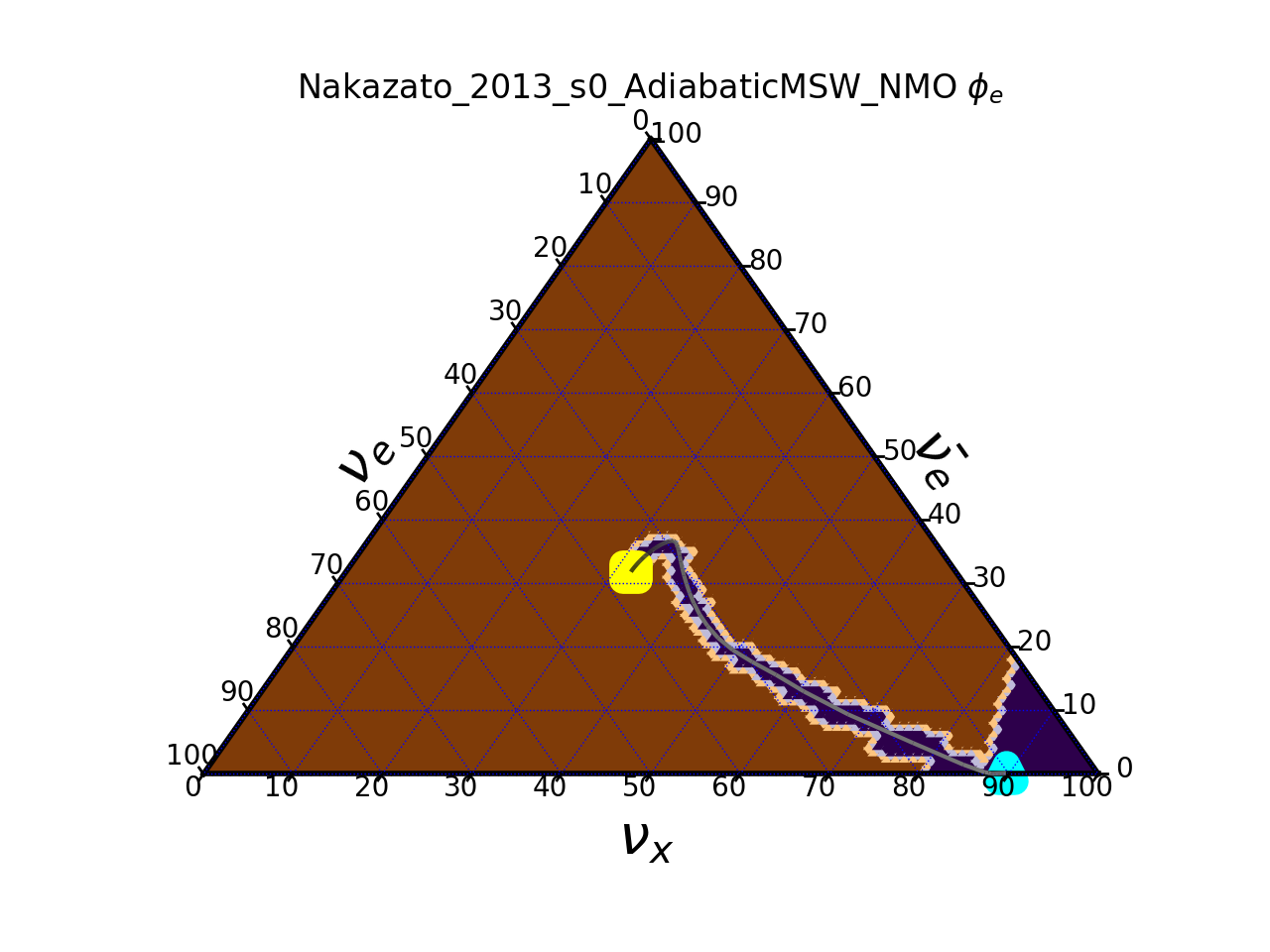} \\
     \hline \\
     \includegraphics[scale=\imotelltalefigscale,trim={0 0 0 1.39cm},clip]{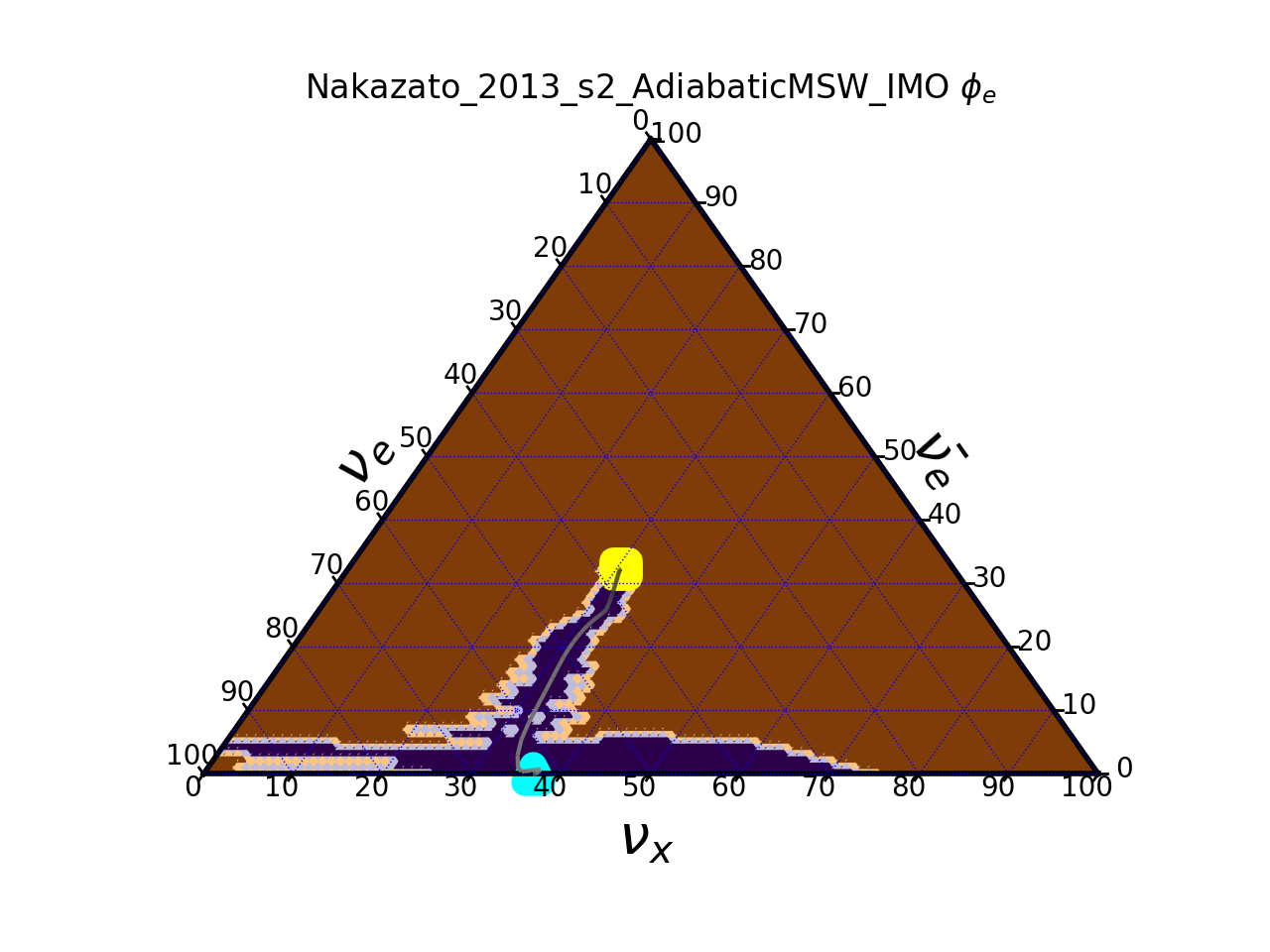} & \includegraphics[scale=\imotelltalefigscale,trim={0 0 0 1.39cm},clip]{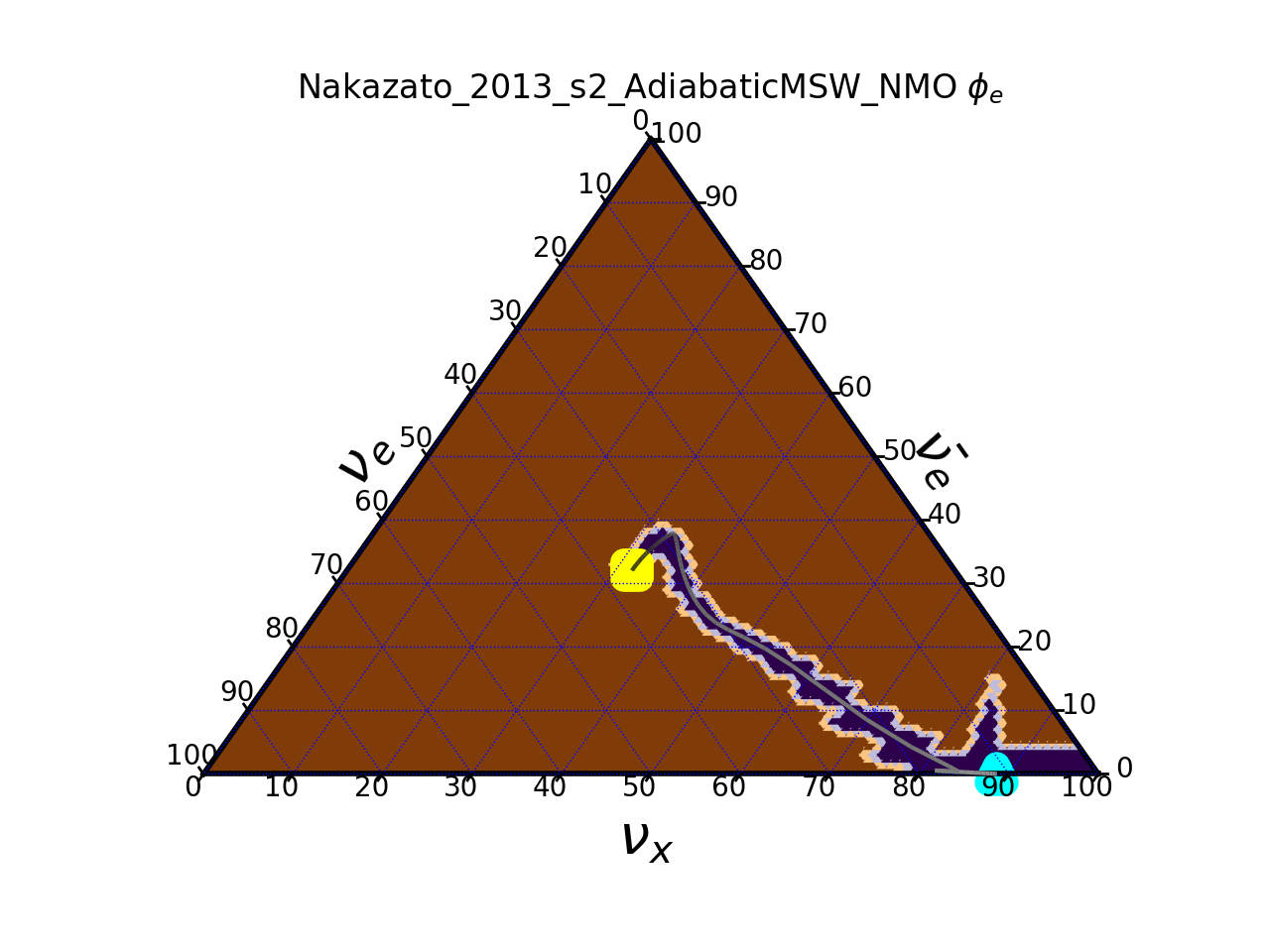} \\
     \hline \\
\end{tabular}

\caption{Unfolded trajectories for Nakazato 2013. By row: model index 0, 2 (see Appendix~\ref{sec:model_descriptions_detailed}). For effective visualization, the $\nu_x$ flavor was divided by six before the fraction was computed. Time bins are logarithmically-separated. Unfolded flux is cumulatively summed over time. The regions are colored according to the prescription in Appendix~\ref{sec:estimated_uncertainties}.}
\label{fig:unfolding_s0_s1}
\end{figure*}

\begin{figure*}

\begin{tabular}{c|c}
     IMO (Nakazato 2013) & NMO (Nakazato 2013) \\
     \hline \hline \\
     \includegraphics[scale=\imotelltalefigscale,trim={0 0 0 1.39cm},clip]{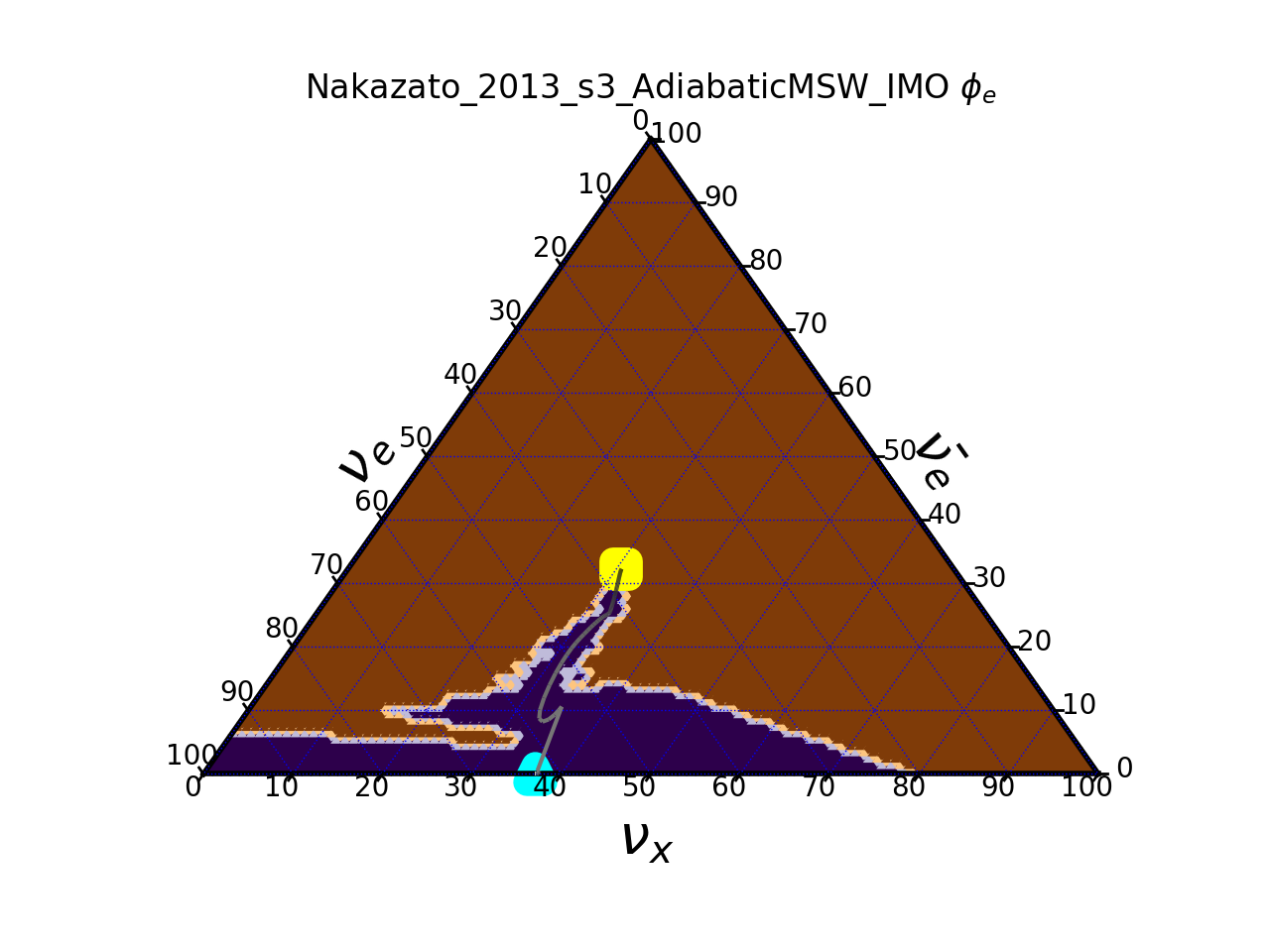} & \includegraphics[scale=\imotelltalefigscale,trim={0 0 0 1.39cm},clip]{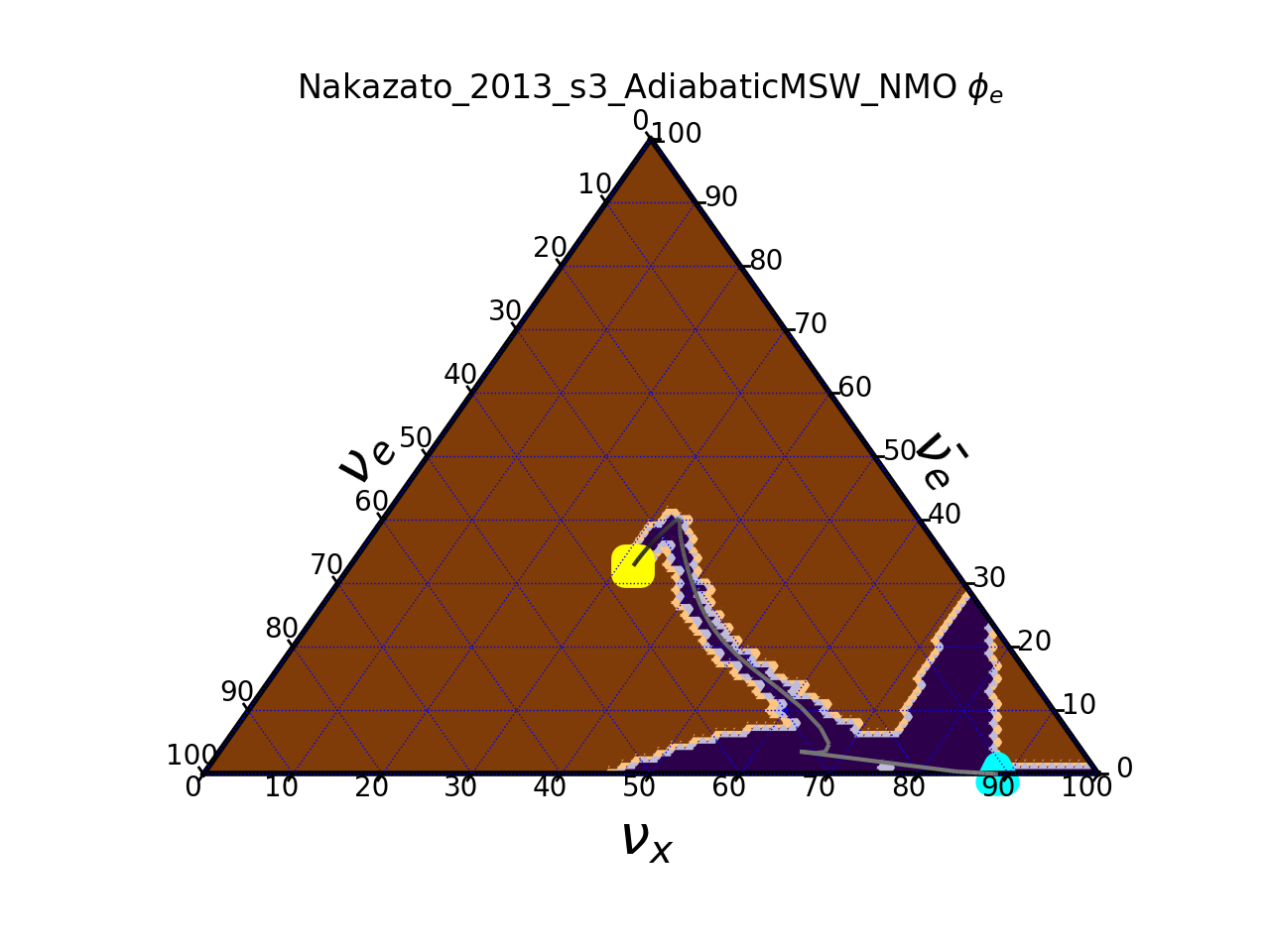} \\
     \hline \\
     \includegraphics[scale=\imotelltalefigscale,trim={0 0 0 1.39cm},clip]{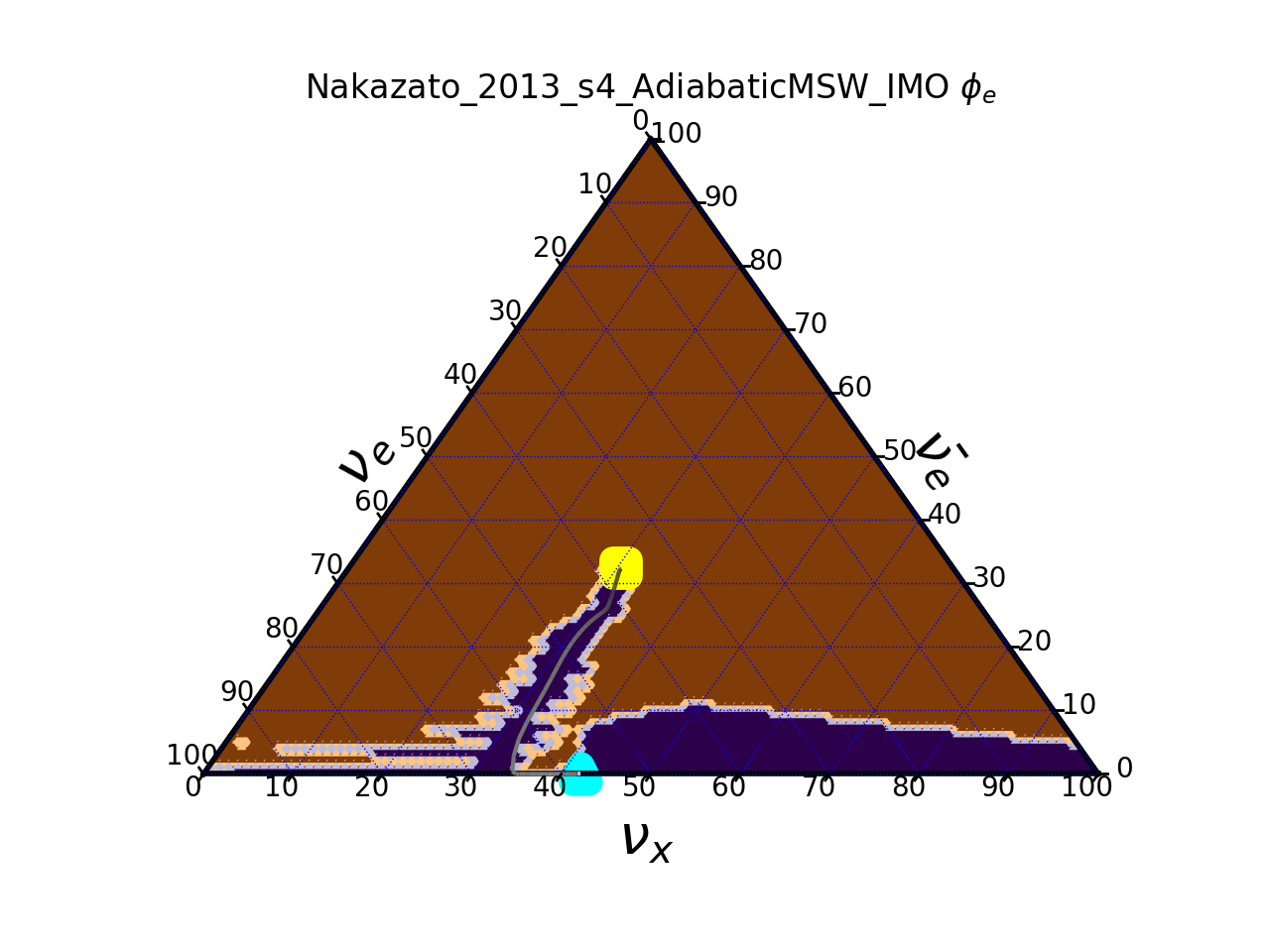} & \includegraphics[scale=\imotelltalefigscale,trim={0 0 0 1.39cm},clip]{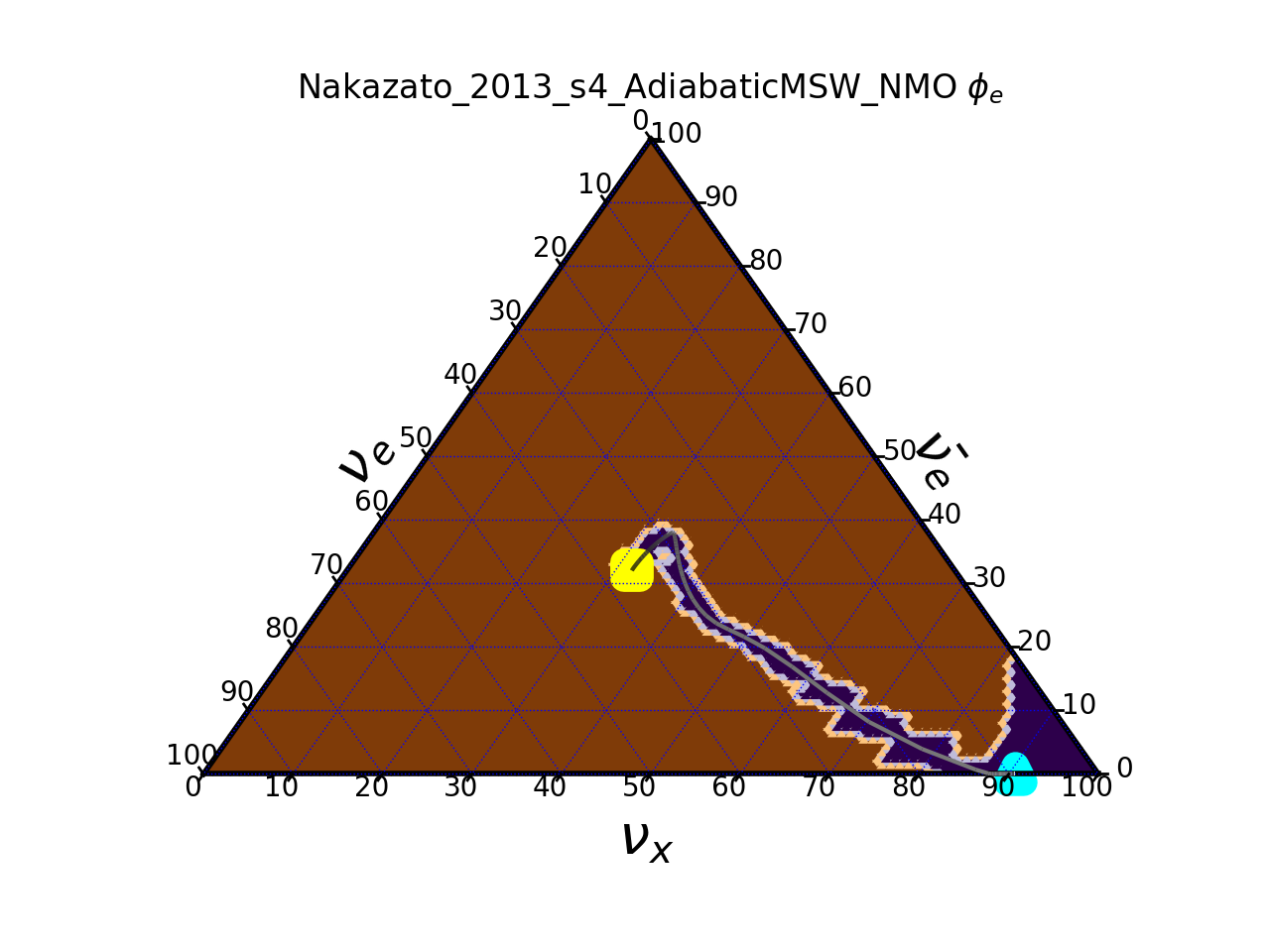}
\end{tabular}

\caption{Unfolded trajectories for Nakazato 2013, plotted in the same way as Fig.~\ref{fig:unfolding_s0_s1}. By row: model index 3, 4 (see Appendix~\ref{sec:model_descriptions_detailed}).}
\label{fig:all_tab:unfolding_s3_s4}
\end{figure*}

\section{Unfolded Fluxes \label{sec:unfolding}}

The number of neutrino-target interaction events per time bin in a detector is given by, 
\begin{equation}
    N=N_t\int_0^\infty \int_{\Delta t_{\rm bin}} \frac{d^2\phi(E_\nu,t)}{dE_\nu dt} \sigma(E_\nu) dt dE_\nu,
\end{equation}

where $N_t$ is the number of targets in the detector, $\sigma(E_\nu)$ is the cross section as a function of neutrino energy for the relevant interaction, and $\frac{d^2\phi(E_\nu,t)}{dE_\nu dt}$  is the neutrino flux from the supernova, differential in energy and time.    The total fluence in the time bin is $\phi_{\rm tot} = \int_0^\infty \int_{\Delta t_{\rm bin}} \frac{d^2\phi(E_\nu,t)}{dt dE_\nu} dE_\nu dt$, and we define the fluence-averaged cross section for the time bin as 
$\langle \sigma \rangle = \int_0^\infty \int_{\Delta t_{\rm bin}} \frac{d^2\phi(E_\nu,t)}{dE_\nu dt} \sigma(E_\nu) dt dE_\nu/\phi_{\rm tot}$.

We can write $N_{t}=\frac{M N_A}{A}$ as the total number of targets in the detector, where $N_A$ is Avogadro's number, $A$ is the atomic mass of the target, and $M$ is the mass of active detector material. The assumed $N_t$ values for the relevant targets for each detector are given in Tab.~\ref{table:Nt}.

\renewcommand{\arraystretch}{1.5}
\begin{table}
\begin{center}
    \begin{tabular}{|| c c c c ||}
    \hline
    Detector & Proxy & Target & $N_t/N_A$ \\
    \hline
    Water 200~kt & $\overline{\nu}_e$ & p & $\frac{(200~{\rm kt})*2}{18~\rm u}$ \\
    Argon 40~kt & ${\nu_e}$ & Ar & $\frac{(40~{\rm kt})}{39.9~\rm u}$\\
    Scint 20~kt & ${\nu_x}$ & $^{12}$C & $\frac{(20~{\rm kt})}{12~\rm u}$ \\
    \hline
    \end{tabular}
\end{center}

\caption{Detector target assumptions for each proxy channel.}
    \label{table:Nt}
\end{table}
\renewcommand{\arraystretch}{1}

For a measured $N$ events in a given time bin, if we know $\langle \sigma \rangle$, we can estimate the fluence in the bin from $\phi_{\rm est} = \frac{N}{N_t \langle \sigma \rangle}$.   Even assuming no uncertainty on $\sigma(E_\nu)$ and $N_t$, in order to estimate $\langle \sigma \rangle$, the standard approach would be to make use of the differential observed energy spectrum in a spectral unfolding or fit of some kind, possibly making use of information detailed detector response, as well as potentially assuming an expected functional form of the differential flux in both time and energy. We are neglecting detector response and associated uncertainties and backgrounds; these effects may or may not be significant for a given channel and detector.  Here, for simplicity, we assume that $\langle \sigma \rangle$ will be reasonably estimated  from such a fit and use the model's truth parameters to estimate it for each time bin.  We incorporate only statistical uncertainties on the counts in each time bin in our uncertainty estimate.  We are fully aware that additional uncertainties in detector response and cross section may affect a realistic answer, but here we are only exploring big-picture patterns.

\paragraph{Simple-Unfolded Ternary Diagram Example}

We next create ternary diagrams from the simple-unfolded fluxes, which should be an approximation of the original truth fluxes, with discrepancies from our simplified unfolding due only to variation over time bins and unaccounted-for detector response.  We then draw error regions reflecting the statistical uncertainty of the measured counts. The counts from each time bin are cumulatively summed, which shrinks the error region along the track. For clear visualization, the individual time bins are not shown in ternary space. Instead, the bins are connected together to form a track. This track is shown as a black and white gradient, where white corresponds to the earliest time point in the model and black corresponds to the end time point of the model. The start and end of the time evolution of the model are denoted by a cyan and yellow triangle, respectively. The confidence region from the detector statistics is overlaid using the approach outlined in App.~\ref{sec:estimated_uncertainties}. If the point is found to be in a 1-$\sigma$ region, it is colored purple; otherwise, it is colored brown. The white shows the boundary between the confidence region and the rest of the distribution.
We show unfolded ternary results from Nakazato 2013 \cite{nakazato_supernova_2013} in Fig.~\ref{fig:unfolding_s0_s1}.

\paragraph{Simple Unfolding Results}

From the collection of models SNEWPY supports, we also show results from additional Nakazato 2013 models,  Warren 2020 \cite{warren_constraining_2020}, and Zha 2021 \cite{zha_progenitor_2021}. We assume that the progenitor is at 10~kpc, and we assume adiabatic MSW flavor transitions.  The ``best channel" configuration is used (see Tab.~\ref{tab:best_channel_proxy_configuration}).

\begin{figure*}
\begin{tabular}{c|c}
     IMO (Warren 2020) & NMO (Warren 2020) \\
     \hline \hline \\
     \includegraphics[scale=\imotelltalefigscale,trim={0 0 0 1.39cm},clip]{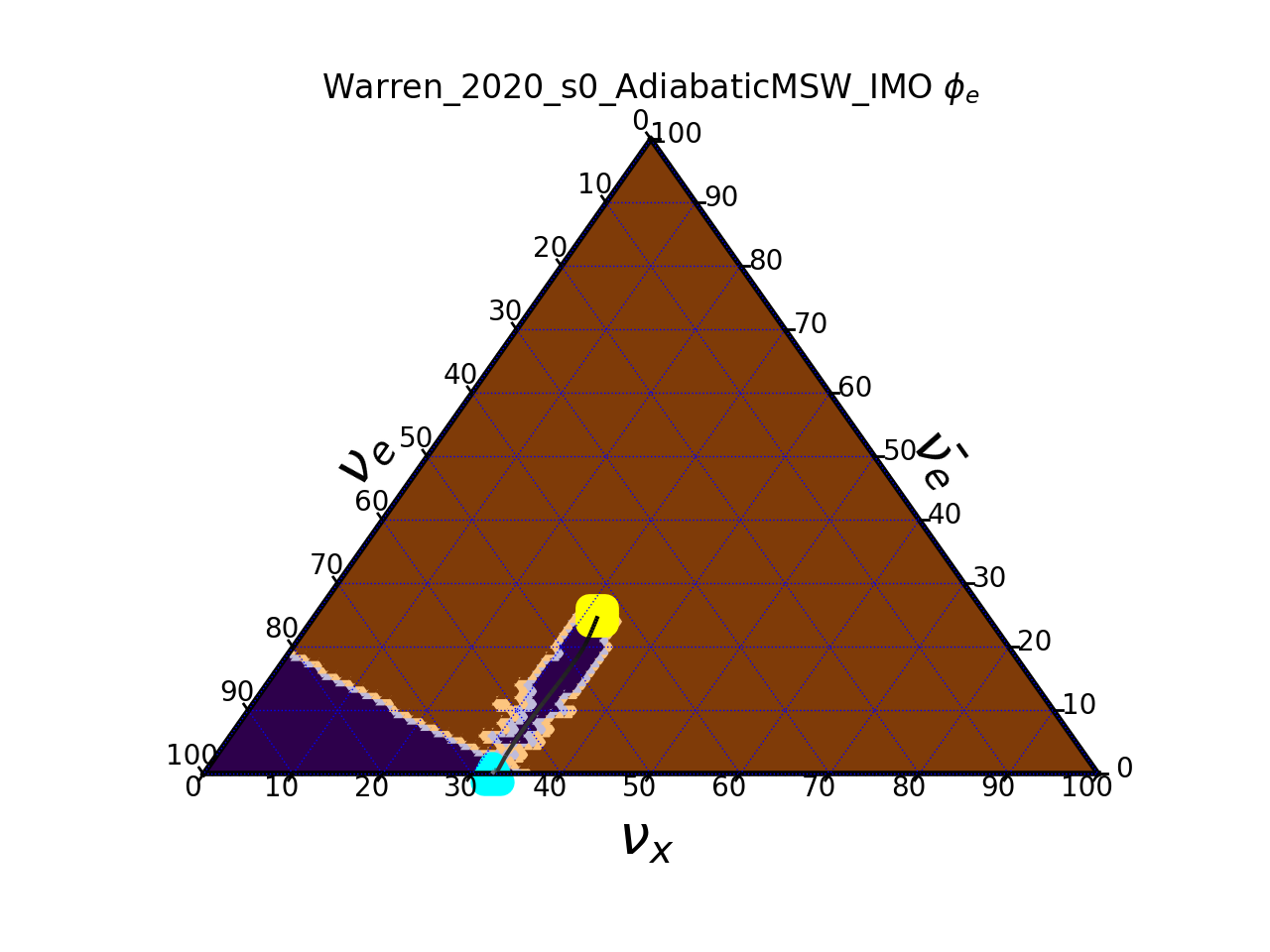} & \includegraphics[scale=\imotelltalefigscale,trim={0 0 0 1.39cm},clip]{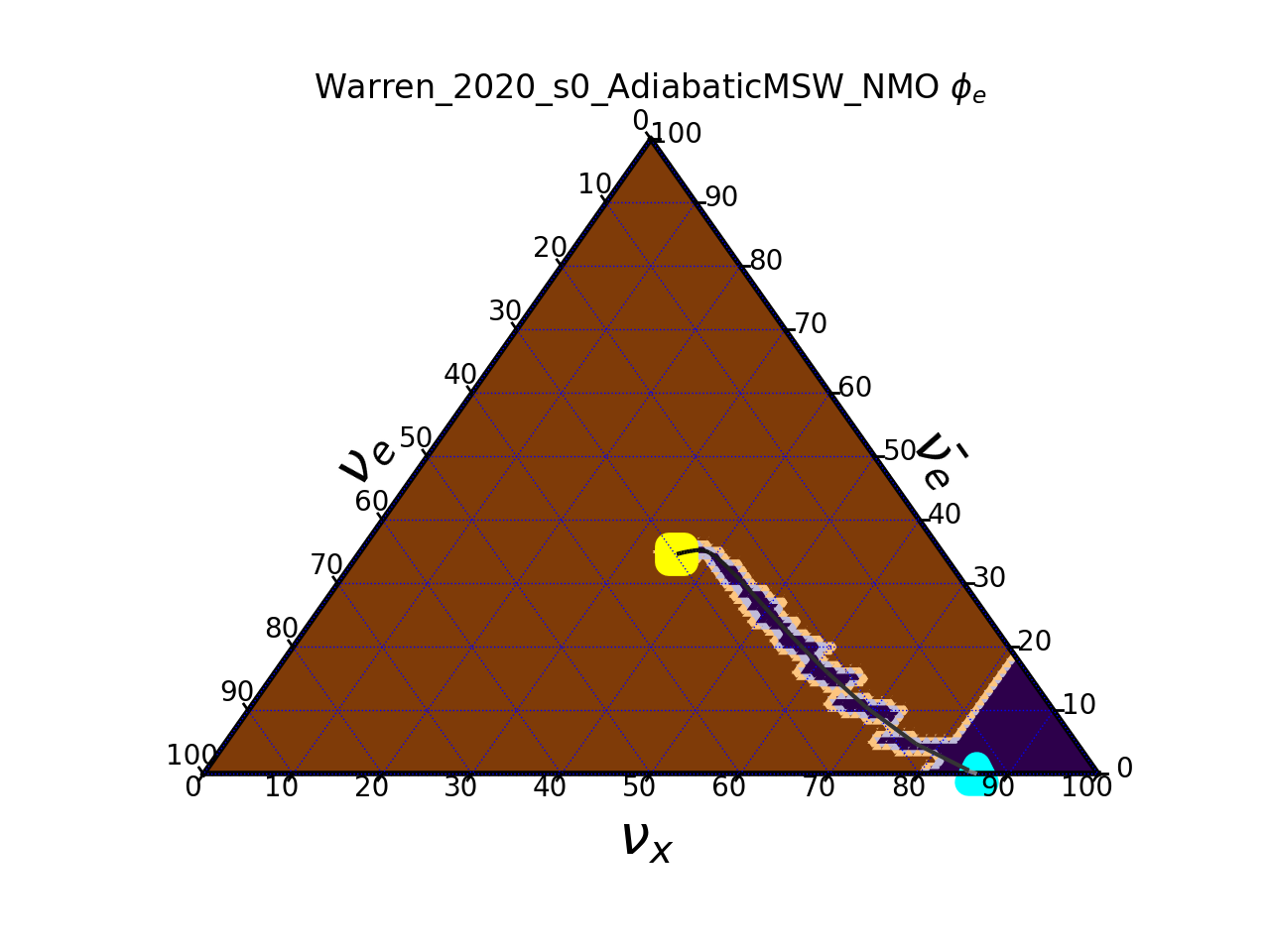} \\
\end{tabular}
\caption{Unfolded trajectories for Warren 2020~\cite{warren_constraining_2020} (see Appendix~\ref{sec:model_descriptions_detailed}), plotted in the same way as Fig.~\ref{fig:unfolding_s0_s1}.}
\label{fig:warren_s0}
\end{figure*}

The results for the ternary diagrams for the unfolded fluxes were more illuminating than those for the simple counts. For example, the ternary diagrams for different models within the Nakazato family (shown in Figs.~\ref{fig:unfolding_s0_s1} and~\ref{fig:all_tab:unfolding_s3_s4}) illustrate that the IMO and NMO regions occupy different parts of the space up to some amount of overlap in error region (1-sigma). Furthermore, the mean trajectories (shown in black and white) all seem to share the same general shape for IMO vs. NMO.

From these visualizations, we find that the NMO and IMO statistically tend to occupy different regions in ternary space. The IMO case tends to have a trajectory with a slope moving from the lower-middle portion of the space towards the center, whereas NMO trajectories tend to slope from the bottom-right portion to the center.  A generally consistent picture holds within a single model family.

\section{Survey of Model Families}\label{sec:models}
\begin{figure}[b]
    \centering
    \includegraphics[width=\linewidth,trim={0 0 0 1.39cm},clip]{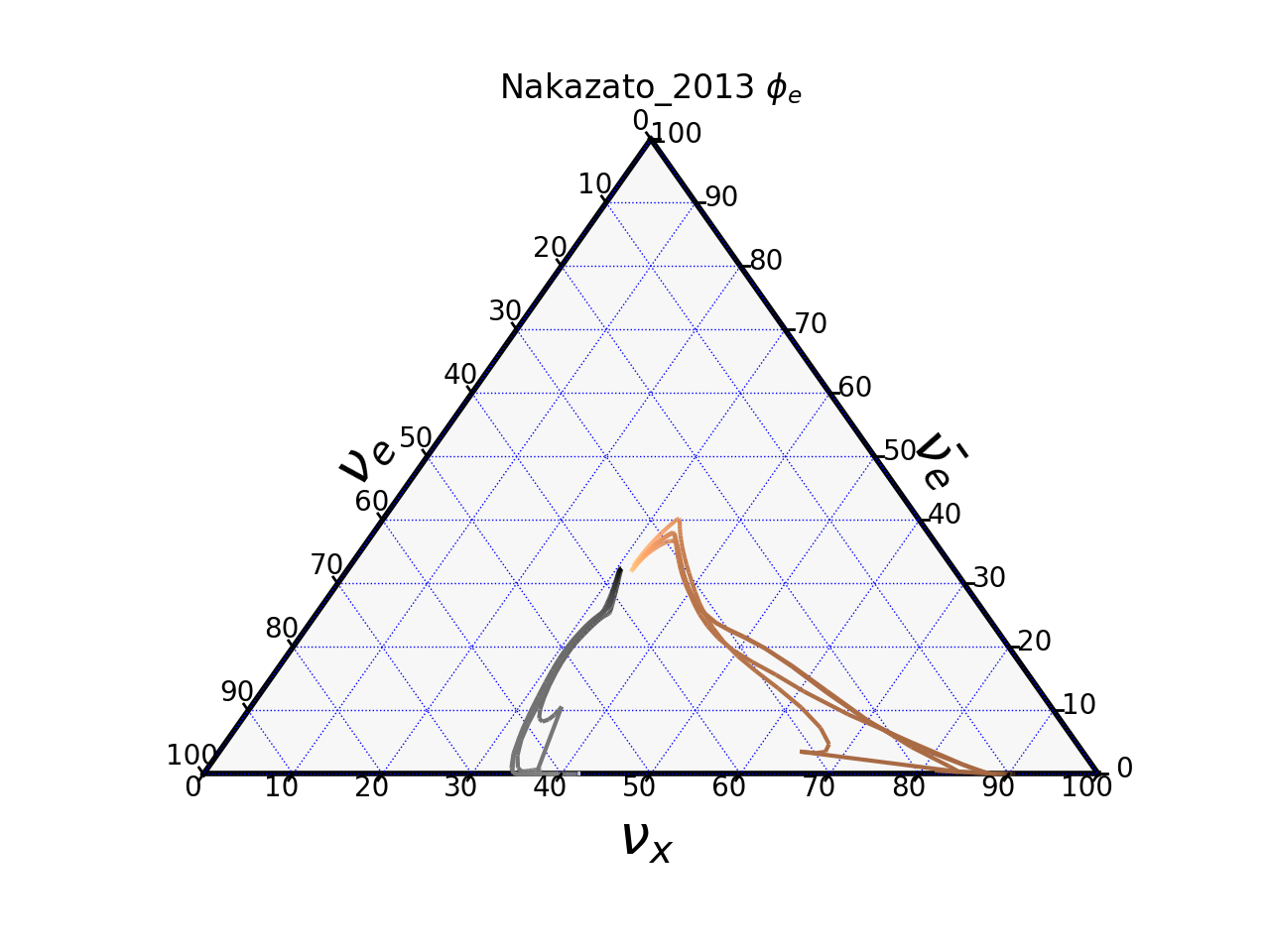}
    \caption{Superimposed cumulatively summed unfolded flux for four Nakazato~\cite{nakazato_supernova_2013} models. For effective visualization, the $\nu_x$ flavor was divided by six before the fraction was computed. The black track is for IMO and the copper track is for NMO. For the copper track, the darker color corresponds to the earlier times. For the gray track, the darker color corresponds to the later times.}
    \label{fig:superimposed-nakazato}
\end{figure}

\begin{figure}
    \centering
    \includegraphics[width=\linewidth,trim={0 0 0 1.39cm},clip]{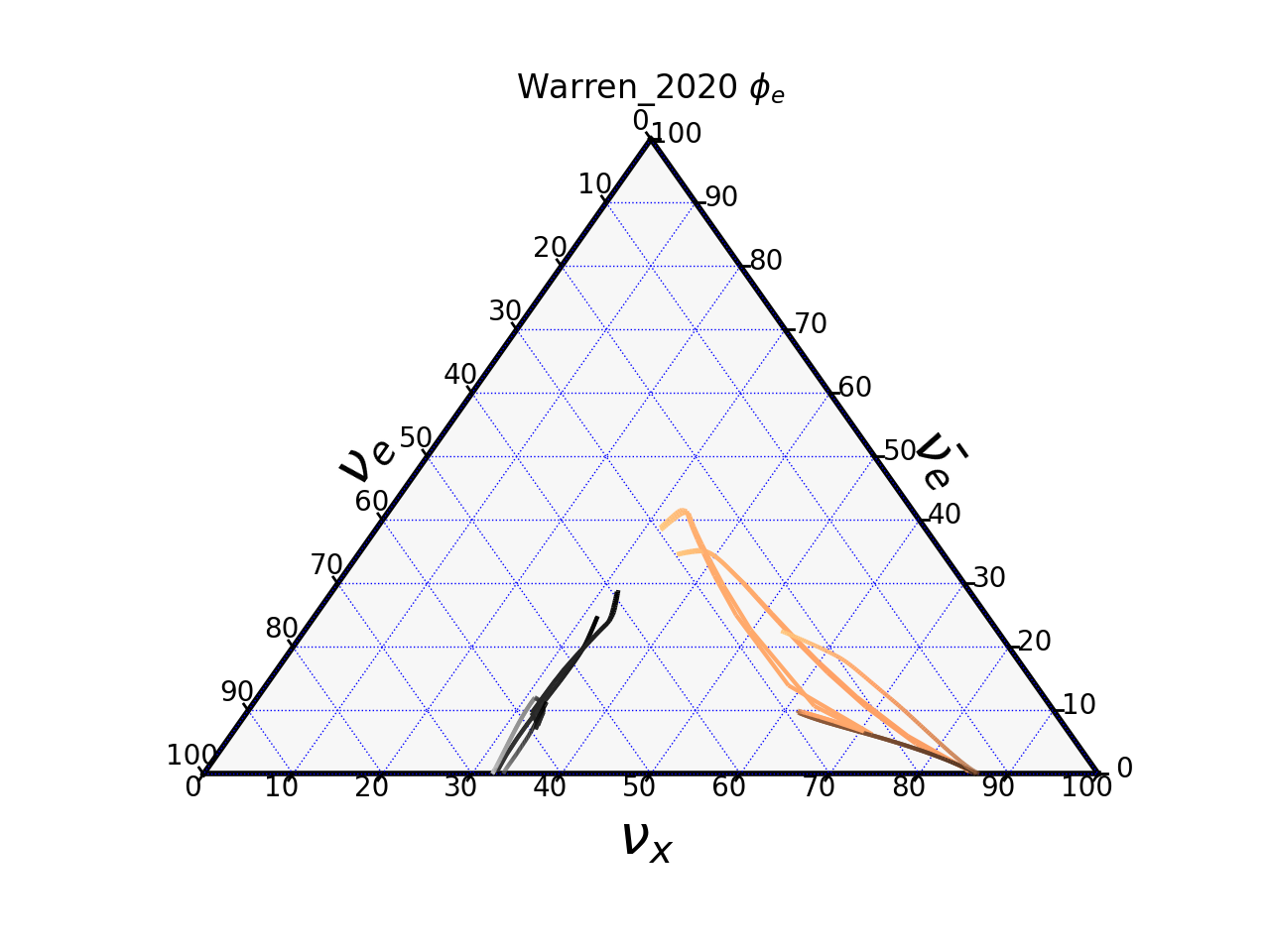}
    \caption{Superimposed cumulatively summed unfolded flux for five Warren~\cite{warren_constraining_2020} models, plotted as in Fig.~\ref{fig:superimposed-nakazato}.}
    \label{fig:superimposed-warren}
\end{figure}

\begin{figure}
    \centering
    \includegraphics[width=\linewidth,trim={0 0 0 1.39cm},clip]{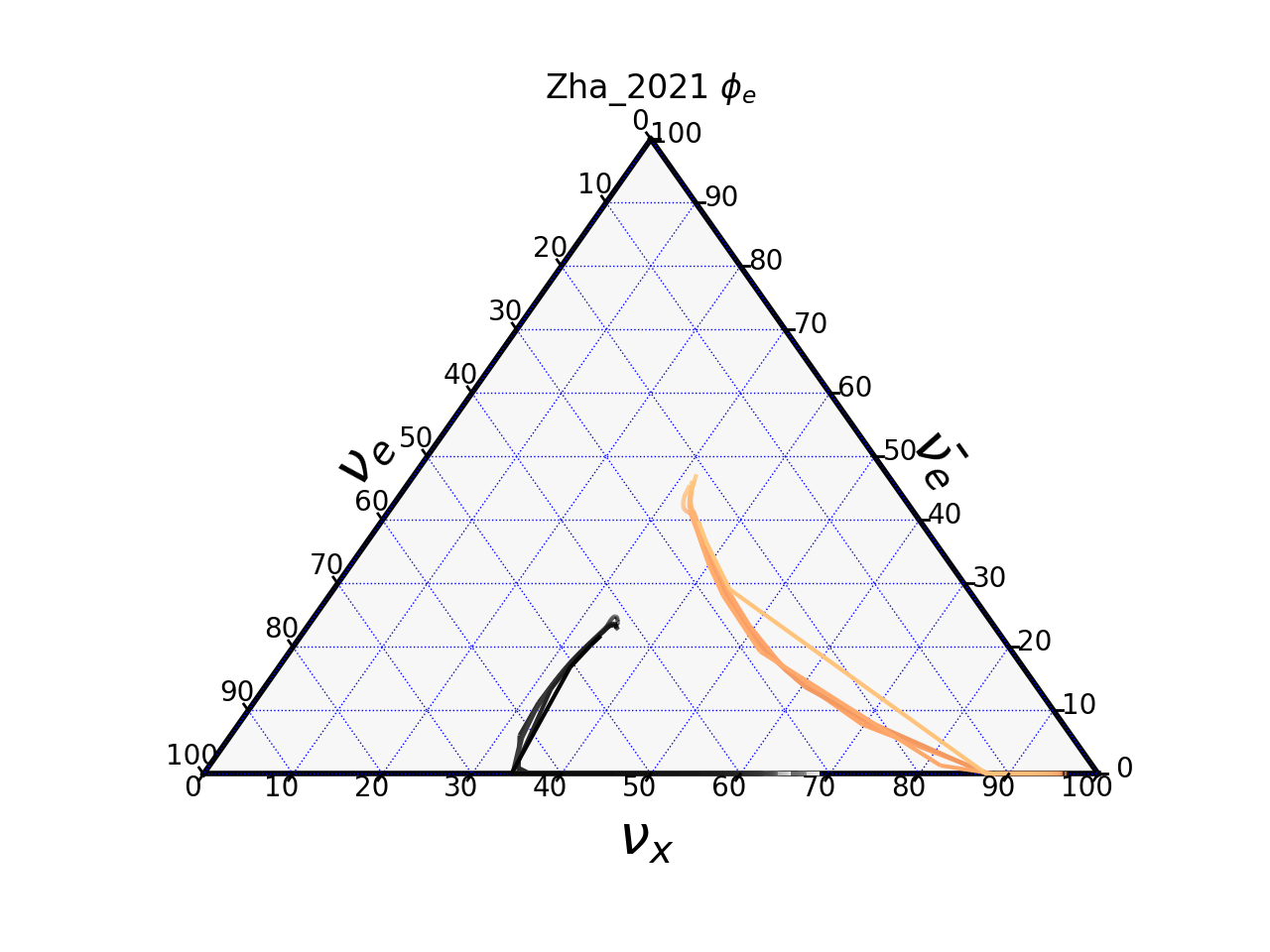}
    \caption{Superimposed cumulatively summed unfolded flux for five Zha~\cite{zha_progenitor_2021} models, plotted as in Fig.~\ref{fig:superimposed-nakazato}.}
    \label{fig:superimposed-zha}
\end{figure}

The same visualization technique was applied for other model families, each of which makes different physical assumptions (see App.~\ref{sec:model_descriptions_detailed}). A set of the trajectories from models from each family are superimposed to compare: Figs. \ref{fig:superimposed-nakazato}, \ref{fig:superimposed-warren}, and \ref{fig:superimposed-zha} show the results of computing the mean IMO/NMO unfolded trajectories for different models. The mean IMO trajectories occupy a similar space as the Nakazato model; the same may be said for the mean NMO trajectories. To illustrate the differences in 1-sigma confidence regions between modelers, we show Warren model 0 in Fig.~\ref{fig:warren_s0}.

From these results, strictly non-overlapping 1-sigma confidence regions were not observed between the NMO and IMO trajectories either within the same model family nor between different model families.  However, the regions tended to occupy different regions in ternary space for the NMO/IMO, even for different model families. One can qualitatively discriminate between the left-hand and right-hand sides of the ternary diagram for IMO and NMO, respectively.

\section{Discussion and Conclusion}\label{sec:discussion}
In this paper, we motivate and outline an approach for using ternary diagrams to visualize the flavor evolution of the neutrino flux from core-collapse supernovae to gain insight into the neutrino mass ordering problem. Analyzing the ternary representations of the supernova flux models from several modelers revealed a possible robust qualitative discriminant between NMO and IMO. The ternary representation of raw detector counts for various models was less illuminating. However, by making use of a simplified unfolding process to approximate the truth flux from detector counts, we found that the flavor trajectories tended to occupy statistically distinct regions of ternary space even over a range of model families. 

We expect that further work that makes use of more sophisticated unfolding will improve understanding of the discrimination potential. For example, one could incorporate uncertainties on cross sections and detector response and take into account the subdominant interaction channels.  Such studies will identify targets for uncertainty reduction.  Ternary track visualization may also aid in the identification of time- and flavor-dependent features over the course of the neutrino burst for assumptions beyond simple MSW transformation models, including potentially collective effects (e.g., Ref.~\cite{Capanema:2024hdm}) or signatures of beyond-the-standard-model physics (e.g., Ref.~\cite{Das:2025zts}.)

\section{Acknowledgments}

The authors are grateful to to Rishi Gundakaram for contributions to the initial versions of some of the software. We thank also Baran Bodur, Janina Hakenm\"uller, Daniel Pershey, Mark Kruse, Leslie Collins, and John Mercer for input and support. This work was supported by t he National Science Foundation, the Society of Physics Students, the Duke Undergraduate Research Support Office, and the Duke Physics Department.

\bibliography{apssamp}

\appendix

\section{Estimated Uncertainties in Ternary Space \label{sec:estimated_uncertainties}}

We assume only uncertainty due to statistics of observed counts at 10~kpc and therefore that the fractional uncertainty of the inferred flux scales as $\sqrt{N}$.
In order to paint an estimated error region in ternary space for $\phi_{\rm est}$, the error for $\phi_{\rm est}$ is propagated using the following prescription.  Let $f_A$ and $f_B$ be two ternary fractions. For counts A, B, and C, the fractions are:
\begin{equation}
    f_A = \frac{A}{A+B+C},~f_B = \frac{B}{A+B+C}
    \label{eqn:fractionaltform}
\end{equation}
$f_C$ is known from $f_A + f_B + f_C = 1$. Assume each of these quantities has an associated Poisson uncertainty and is statistically independent, i.e., $A\pm \sqrt{A}$, $B\pm\sqrt{B}$, $C\pm\sqrt{C}$. We are interested in finding the propagated uncertainties on the fractions, $f_A\pm\delta f_A$ and $f_B\pm \delta f_B$.  The propagated error matrix is given by
\begin{equation}
    \begin{split}
        \boldsymbol{\Sigma_f} =\boldsymbol{D}^T\boldsymbol{\Sigma_N}\boldsymbol{D}
        &= \boldsymbol{D}^T\begin{bmatrix}
            A & 0\\
            0 & B
        \end{bmatrix}\boldsymbol{D}\\
        &=
\begin{bmatrix}
 \frac{A (B+C)}{(A+B+C)^3} & -\frac{A B}{(A+B+C)^3}\\
 -\frac{A B}{(A+B+C)^3} & \frac{B (A+C)}{(A+B+C)^3}
\end{bmatrix}
    \end{split}
\end{equation}
where the derivatives matrix, $\boldsymbol{D}$, is
\begin{equation}
    \begin{bmatrix}
        \frac{\partial f_A}{\partial A} & \frac{\partial f_A}{\partial B} \\
        \frac{\partial f_B}{\partial A} & \frac{\partial f_B}{\partial B}
    \end{bmatrix}
\end{equation}
We then treat the PDF (probability density function) for the propagated quantity $f_{A,B}$ as a multi-variate Gaussian distribution.
We follow the same process for finding the uncertainty on $\phi$ using the forms of $f_A$, $f_B$  \footnote{$f_C$ is left out due to full correlation between $f_C$ and $f_A$ and $f_B$.}. We define fractions $f_{\phi_x}=\frac{\phi_x}{\phi_x+\phi_y+\phi_z}$, $f_{\phi_y}=\frac{\phi_y}{\phi_x+\phi_y+\phi_z}$, and $f_{\phi_z}=\frac{\phi_z}{\phi_x+\phi_y+\phi_z}$. We take $\vec{\phi} = (\phi_x,\phi_y,\phi_z)$ be the $\phi_{\rm est}$ for the three flavors in a time bin, and $\vec{N} = (N_x,N_y,N_z)$ are the respective $N_{\rm det}$ counts used to unfold the $\phi$ values. The error matrix for $\phi$ is then:
\begin{equation}
    \boldsymbol{\Sigma_\phi}=
    \begin{bmatrix}
        \sigma_{\phi_x}^2 & \text{cov}(\phi_x,\phi_y) \\
        \text{cov}(\phi_x,\phi_y) & \sigma_{\phi_y}^2
    \end{bmatrix}=
    \begin{bmatrix}
        \frac{\phi_x^2}{N_x} & 0 \\
        0 & \frac{\phi_y^2}{N_y}
    \end{bmatrix}
    \label{eqn:phi_error_matrix}
\end{equation}
We then have using Eq.~\ref{eqn:phi_error_matrix} and the derivatives matrix,
\begin{equation}
    \begin{split}
        \boldsymbol{D}=\begin{bmatrix}
            \frac{\partial f_{\phi_x}}{\partial\phi_x}&\frac{\partial f_{\phi_x}}{\partial\phi_y}\\
            \frac{\partial f_{\phi_y}}{\partial\phi_x}&\frac{\partial f_{\phi_y}}{\partial\phi_y}
        \end{bmatrix}
    \end{split}
\end{equation}
to arrive at the final propagated error matrix for the $f_\phi$ fractions:

\onecolumngrid

\begin{widetext}
\begin{equation}
\begin{split}
   \boldsymbol{\Sigma}_{\phi_f}= &\boldsymbol{D}^T\boldsymbol{\Sigma_\phi}\boldsymbol{D} =\\
    &\begin{bmatrix}
    \frac{\phi _x^2 \left(N_x \left(N_z \phi _y^2+N_y \phi _z^2\right)+N_y N_z \left(\phi _y+\phi _z\right){}^2\right)}{N_x N_y N_z \left(\phi _x+\phi _y+\phi _z\right){}^4} & -\frac{\phi _x \phi _y \left(-N_x N_y \phi _z^2+N_x N_z \phi _y \left(\phi _x+\phi _z\right)+N_y N_z \phi _x \left(\phi _y+\phi _z\right)\right)}{N_x N_y N_z \left(\phi _x+\phi _y+\phi _z\right){}^4} \\
    -\frac{\phi _x \phi _y \left(-N_x N_y \phi _z^2+N_x N_z \phi _y \left(\phi _x+\phi _z\right)+N_y N_z \phi _x \left(\phi _y+\phi _z\right)\right)}{N_x N_y N_z \left(\phi _x+\phi _y+\phi _z\right){}^4} & \frac{\phi _y^2 \left(n_y N_z \phi _x^2+N_x \left(N_z \left(\phi _x+\phi _z\right){}^2+N_y \phi _z^2\right)\right)}{N_x N_y N_z \left(\phi _x+\phi _y+\phi _z\right){}^4}
    \end{bmatrix}
\end{split}
    \label{eqn:full_propagated_error}
\end{equation}
\end{widetext}
\twocolumngrid

From this propagated error matrix, we construct a multi-variate Gaussian PDF for the ternary fractions.  For each point of the ternary trajectory, we color points in the ternary diagram purple that are within a 68\% C.L. region.   The ternary visualization for the full trajectory has points colored purple if within the C.L. region of any trajectory point.

\section{Model Descriptions \label{sec:model_descriptions_detailed}}
In this section, we list a few details of the models used in this analysis, particularly the ones that are used to produce Figs. \ref{fig:superimposed-nakazato}, \ref{fig:superimposed-warren}, and \ref{fig:superimposed-zha}.

The Nakazato 2013 model family~\cite{nakazato_supernova_2013} is a collection of core-collapse supernova simulations in which the authors vary the progenitor masses over a range of $13 M_\odot-50M_\odot$ and metallicities $Z\in\{0.02,0.004\}$.

\begin{center}
    \textbf{Nakazato 2013 \cite{nakazato_supernova_2013}}
    \begin{tabular}{|| c c c ||}
    \hline
    Index & Prog. Mass & $Z$ \\
    \hline
    0 & $20~M_\odot$ & $0.004$ \\
    2 & $13~M_\odot$ & $0.004$ \\
    3 & $50~M_\odot$ & $0.004$ \\
    4 & $13~M_\odot$ & $0.02$ \\
    \hline
    \end{tabular}
\end{center}

The Warren 2020~\cite{warren_constraining_2020} model family contains several core-collapse supernova simulations that focus on the one-dimensional hydrodynamical physics of the star. The simulations vary in progenitor mass from $9 M_\odot-120 M_\odot$ and in a turbulence mixing parameter used to describe the hydrodynamics.

\begin{center}
    \textbf{Warren 2020 \cite{warren_constraining_2020}}
    \begin{tabular}{|| c c c ||}
    \hline
    Index & Prog. Mass & Turb. Mixing \\
    \hline
    0 & $10~M_\odot$ & $1.23$ \\
    1 & $25~M_\odot$ & $1.23$ \\
    2 & $10~M_\odot$ & $1.27$ \\
    3 & $25~M_\odot$ & $1.27$ \\
    4 & $55~M_\odot$ & $1.25$ \\
    \hline
    \end{tabular}
\end{center}

The Zha 2021~\cite{zha_progenitor_2021} model family focuses on simulations of the effects of hadron-quark phase transitions in core-collapse supernovae. The simulations feature a specific equation of state and vary in progenitor mass.
\begin{center}
    \textbf{Zha 2021 \cite{zha_progenitor_2021}} \\
    \begin{tabular}{|| c c ||}
    \hline
    Index & Prog. Mass\\
    \hline
    0 & $16~M_\odot$ \\
    1 & $17~M_\odot$ \\
    2 & $18~M_\odot$ \\
    3 & $19~M_\odot$ \\
    4 & $20~M_\odot$ \\
    \hline
    \end{tabular}
\end{center}

\end{document}